\documentclass[]{aastex631}

\usepackage{color, colortbl}
\usepackage{amsmath}
\usepackage{txfonts}
\usepackage{mathtools}
\usepackage{units}
\usepackage[Symbolsmallscale]{upgreek}
\usepackage{xargs}
\usepackage{xspace}

\graphicspath{{Images/}{Images/inspections/}} 

\definecolor{Gray}{gray}{0.7}
\definecolor{lightGray}{gray}{0.85}
\definecolor{darkblue}{rgb}{0.0,0.0,0.4}
\definecolor{darkgreen}{rgb}{0.0,0.4,0.0}
\definecolor{darkred}{rgb}{0.5,0.0,0.0}
\definecolor{grey}{rgb}{0.4,0.4,0.6}

\newcommand{\Kt}[0]{\citetalias{2018A&A...620A.175K}\xspace}

\newcommand{\eDR}[0]{\textit{Gaia}\,eDR3\xspace}
\newcommand{\DRtwo}[0]{\textit{Gaia}\,DR2\xspace}
\newcommand{\Gaia}[0]{\textit{Gaia}\xspace}
\newcommand{\mas}[0]{\,\mathrm{mas}}
\newcommand{\MAG}[0]{\,\mathrm{mag}}
\newcommand{\mmag}[0]{\,\mathrm{mmag}}
\newcommand{\yr}[0]{\,\mathrm{yr}}
\newcommand{\Sou}[0]{\mathrm{Sou}\_}
\newcommand{\mpy}[0]{\,\mathrm{mas/yr}}
\newcommand{\Msun}[0]{\,\mathrm{M_{\odot}}}
\defcitealias{2018A&A...620A.175K}{K18b}

\newcommand{\urlpython}[0]{\url{https://github.com/jkluter/amlensing}\xspace}
\newcommand{\urlresult}[0]{\url{https://dc.g-vo.org/amlensing/q3/q/form}\xspace}
\newcommand{\urltab}[0]{\url{https://dc.g-vo.org/tap}\xspace}

\shorttitle{Astrometric-Microlensing Events from {\it Gaia} eDR3}
\shortauthors{Kl\"{u}ter et al.}
\received{13 Sep. 2021}
\accepted{17 Dec. 2021}
\submitjournal{AJ}
\begin{document}

\title{Prediction of Astrometric-Microlensing Events from  \eDR Proper Motions
\footnote{The results of the predicted microlensing events is available in electronic form via the GAVO Data Center \urlresult}
\footnote{The source code for this study is made publicly available \urlpython}}
\correspondingauthor{Jonas Kl\"{u}ter}
\email{jklueter1@lsu.edu}

\author{Jonas Kl\"{u}ter}
\affiliation{Department of Physics and Astronomy, Louisiana State University, 202 Nicholson Hall, Baton Rouge, LA 70803 USA}
\affiliation{Zentrum f\"{u}r Astronomie der Universit\"{a}t Heidelberg, Astronomisches Rechen-Institut, M\"{o}nchhofstr. 12-14, 69120 Heidelberg, Germany}

\author{Ulrich Bastian}
\affiliation{Zentrum f\"{u}r Astronomie der Universit\"{a}t Heidelberg, Astronomisches Rechen-Institut, M\"{o}nchhofstr. 12-14, 69120 Heidelberg, Germany}

\author{Markus Demleitner}
\affiliation{Zentrum f\"{u}r Astronomie der Universit\"{a}t Heidelberg, Astronomisches Rechen-Institut, M\"{o}nchhofstr. 12-14, 69120 Heidelberg, Germany}

\author{Joachim Wambsganss}
\affiliation{Zentrum f\"{u}r Astronomie der Universit\"{a}t Heidelberg, Astronomisches Rechen-Institut, M\"{o}nchhofstr. 12-14, 69120 Heidelberg, Germany}
\affiliation{International Space Science Institute, Hallerstr. 6, 3012 Bern, Switzerland}

\begin{abstract}

Astrometric microlensing is a unique tool to measure stellar masses. It allows us to determine the mass of the lensing star with an accuracy of a few per cent. 
In this paper, we update, extend, and refine our predictions of astrometric-microlensing events based on Gaia's early Data release 3 (eDR3).
We selected about 500.000 high-proper-motion stars from Gaia~eDR3 with \(\mu_{tot}>100\mpy\) and searched for background sources close to their paths. 
We applied various selection criteria and cuts in order to exclude spurious sources and co-moving stars.   
By forecasting the future positions of lens and source we determined epoch of and angular separation at closest approach, and determined an expected positional shift and magnification.
Using Gaia~eDR3, we predict 1758 new microlensing events with expected shifts larger than \(0.1\mas\) between the epochs J2010.5 and mid J2066.0. Further we provide more precise information on the angular separation at closest approach for 3084 previously predicted events.
This helps to select better targets for observations, especially for events which occur within the next decade.
Our search lead to the new prediction of an interesting astrometric-microlensing event by the white dwarf Gaia~eDR3-4053455379420641152.
In 2025 it will pass by a \(G=20.25\MAG\) star, which will lead to a positional shift of the major image of \(\delta\theta_{+}=1.2^{+2.0}_{-0.5}\mas\).
  Since the background source is only \(\Delta G=2.45\MAG\) fainter than the lens, also the shift of the combined center of light will be measurable, especially using a near infrared filter, where the background star is brighter than the lens (\(\Delta Ks=-1.1\MAG\)).

\end{abstract}


\section{Introduction}

The mass of a star is one of its most important parameters. It defines its luminosity, temperature, surface gravity, appearance, and evolutionary path. Testing evolutionary and stellar models requires accurate and direct measurements of fundamental stellar parameters. Direct masses are usually derived from double-lined spectroscopic and eclipsing binaries. However, for most of the isolated stars, masses can only be derived indirectly, typically by using the mass-luminosity or mass-radius relations. For the determination of such relations, a set of accurately known masses is required. These are mainly derived from binary stars \citep{1991A&ARv...3...91A, 2010A&ARv..18...67T}. 
However, binary stars and isolated stars may evolve differently. Therefore it is not known how well these empirical relations describe the masses of single stars. For a better understanding of the mass-luminosity relations, direct mass measurements of single stars are important. Besides asteroseismology, which itself is strongly model dependent, gravitational microlensing is the only available tool.
Further, the direct determined mass of white dwarfs provides a unique test sample for comparison with theoretical mass-radius relations and evolutionary cooling tracks of white dwarfs, a first such measurement was achieved by \cite{2017Sci...356.1046S}.

As a sub-area of gravitational lensing, microlensing describes the time-dependent positional deflection (astrometric microlensing) and magnification (photometric microlensing) of a background source (BGS) due to an intervening star (“lens”) \citep{1986ApJ...301..503P,1986ApJ...304....1P}. Using microlensing, it is possible to determine the mass of the lens star with uncertainties in the order of a few per cent, either by detecting finite-source effects in photometric microlensing events and measuring the microlens parallaxes \citep{1992ApJ...392..442G} or by 
observing the positional deflection of astrometric-microlensing events \citep{1991ApJ...371L..63P,1995AcA....45..345P}. 
While the usage of photometric microlensing is slightly dependent on empirical relations to determine the radius of the BGS, astrometric microlensing is completely model independent and depends directly on the mass of the lens and its distance. In the last few years, this was used by \cite{2017Sci...356.1046S} and \cite{2018MNRAS.480..236Z} to determine the masses of Stein 2051b and Proxima Centauri, respectively.  %
In comparison to photometric microlensing, astrometric microlensing has two additional advantages: It can be observed at larger angular separations between the background source and the foreground lens which results furthermore in a longer time-scale for the event of the order of months to years \citep{2000ApJ...534..213D}. The second advantage is the possibility to confidentially predict astrometric-microlensing events for stars with known proper motions \citep{1964MNRAS.128..307R,1995AcA....45..345P}. Further, these precise predictions are also needed for the mass determination to derive the unlensed separation. 

In 2000, \cite{2000ApJ...539..241S} systematically searched for astrometric-microlensing events for the first time, followed by several studies based on non Gaia proper motions \citep{2011A&A...536A..50P, 2012ApJ...749L...6L, 2018MNRAS.475...79H}. Nowadays the most precise predictions make use of astrometric data from the \Gaia satellite \citep{2016A&A...595A...1G}.
Using \Gaia DR1, the first data release, \cite{2018MNRAS.478L..29M}  predicted one event caused by a white dwarf in 2019. With the second \Gaia data release \citep[\DRtwo,][]{2018A&A...616A...1G} about 5700 events were predicted, either by using solely \DRtwo\citep{2018A&A...615L..11K, 2018A&A...618A..44B, 2018AcA....68..183B, 2018A&A...620A.175K, 2019MNRAS.483.4210M} or by combining it with additional catalogues \citep{2018AcA....68..351N,2019MNRAS.487L...7M}, such as Pan-STARRS \citep{2016arXiv161205560C}\footnote{Panoramic Survey Telescope and Rapid Response System, \\ \url{https://panstarrs.stsci.edu}}, or VVV, \citep{2010NewA...15..433M}\footnote{VISTA Variables in the Via Lactea \url{https://vvvsurvey.org}}, respectively.
Additionally, \cite{2018A&A...617A.135M} searched for astrometric-microlensing events with impact parameters of the order of one Einstein radius. Such events will lead to a potentially measurable photometric magnification of the source star.  
They found 30 such events that will occur until the year 2032 which have a \(>\)10\%   probability to have an impact parameter smaller than the Einstein radius.
However, the astrometric precision of \DRtwo (and also of eDR3) is not sufficient to securely predict photometric microlensing events.

 In December 2020, \Gaia published its early Data Release 3 \cite[\eDR,][]{refId6}, which contains updated and more precise astrometric parameters of about two billion stars. 
\cite{McGill2020} showed that events predicted from \DRtwo which occur in the coming few years are contaminated by spurious background stars and by pairs of co-moving stars. 
\eDR includes fewer spurious sources, and a better exclusion of co-moving stars is possible due to the larger fraction of stars with available proper motions. 
Further, the increase in the astrometric precision allows more accurate predictions.

In this paper we update our previous search for astrometric-microlensing events in \citet[][hereafter K18b]{2018A&A...620A.175K} by using  \eDR.
First, in Section~\ref{Section:AML}, we shortly explain astrometric and photometric microlensing. 
In Section~\ref{Section:Search} we explain our search for astrometric-microlensing events, starting with the selection of potential lens and source stars in \ref{subSection:HPMS} and \ref{subSection:BGS},  respectively, and continuing with the determination of approximate masses and Einstein radii in \ref{subSection:Mass}, the forecast of their paths and the detection of close encounters in \ref{subSection:FindClosest}, and the determination of the expected effects and selection of observable events in \ref{subSection:effect}. 
In Section~\ref{Section:Results}, we present the predicted astrometric-microlensing events and compare this list with previous studies. Finally, we draw conclusions in Section~\ref{Section:Conclusions}

\section{Basics of microlensing}
\label{Section:AML}
Microlensing describes the time-dependent photometric magnification and positional change of a background source by a point-like foreground mass (``lens'') passing by.
While the lens is passing the source, two images of the source are created: a bright major image (\(+\)) close to the unlensed source position, and a faint minor image (\(-\)) close to the lens position.
In case of perfect alignment, the background source appears as a so-called Einstein ring. Its radius is given by \citep{1924AN....221..329C,1936Sci....84..506E, 1986ApJ...301..503P}:  
\begin{equation}
\theta_{\rm E} = \sqrt{\frac{4GM_{\rm L}}{c^{2}}\frac{D_{\rm S}-D_{\rm L}}{D_{\rm S}\cdot D_{\rm L}}} = 2.854 \mas \sqrt{\frac{M_{\rm L}}{\Msun} \frac{\varpi_{\rm L} - \varpi_{\rm S}}{1\mas}}
\label{EQ:ThetaE}
\end{equation}
where \(G\) is the gravitational constant, \(c\) is the speed of light, \(M_{L}\) is the mass of the lens, \(D_{\rm L}\), \(D_{\rm S}\) are the distances of  lens and source from the observer, and \(\varpi_{\rm L}\), \(\varpi_{\rm S}\) are the parallaxes of lens and source, respectively. This Einstein radius is often used as an angular scale for the microlensing event, e.g.~to define the (unlensed) scaled angular separation on the sky \(\boldsymbol{u} =\boldsymbol{\theta}/ \theta_{\rm E}\), where \(\boldsymbol{\theta}\) is the two-dimensional unlensed angular separation.

\subsection{Photometric microlensing}
The photometric magnification only depends on this unit-less angular separation \(\boldsymbol{u}\).
The total magnification of the two images is given by \citep{1986ApJ...301..503P}:
\begin{equation}
A = A_{+} + A_{-} = \frac{u^{2}+2}{u\sqrt{u^{2}+4}}, 
\end{equation} 
where \(u = \lvert \boldsymbol{u} \rvert \). 
The apparent magnification of the background source can be reduced due to blending with light from additional unresolved light sources.
For the here predicted events, the most important additional light source is the lens itself.  
With a flux ratio \(f_{\rm LS}\) between lens and source, it can be expressed as \citep{2000ApJ...534..213D}:
\begin{equation}
A_{\rm lum} = \frac{f_{\rm LS} + A}{f_{\rm LS} +1},
\end{equation} 
and in units of magnitude, it is given by
\begin{equation}
 \label{EQ:mag}
\Delta m = 2.5 \cdot\log_{10}\left({\frac{f_{LS}+A}{f_{LS}+1}}\right). 
\end{equation} 
Due to a strong decline  \citep[\(A  \simeq 1+ 2/u^{4}\),][]{1996AcA....46..291P}, a magnification is only observable when the impact parameter is on the order of the Einstein radius. This also means that a photometric microlensing event usually has a short time scale in the order of days to weeks. This can be expressed by the Einstein time \citep{1992ApJ...392..442G}:
\begin{equation}
t_{E}=\frac{\theta_{E}}{\mu_{rel}}, 
\end{equation} 
where \(\mu_{rel}\) is the absolute value of the relative proper motion between lens and source.

\subsection{Astrometric microlensing}

In astrometric microlensing, the signal of interest is the change of the positions of the two images. The positional shift of the major image with respect to the unlensed position of the source is given by:
\begin{equation}
 \label{EQ:shiftp}
\delta\boldsymbol{\theta_{+}} = \frac{  \sqrt{(u^{2}+4)} - u}{2} \cdot \frac{\boldsymbol{u}}{u} \cdot{\theta_{\rm E}}.
\end{equation} 
In the unresolved case, this deflection is reduced due to blending.
First, due to the minor image, which leads to a shift of the centre of light (in respect to the unlensed position of the source),
\begin{equation}
\label{EQ:shiftC}
\delta\boldsymbol{\theta_{c}} = \frac{\boldsymbol{u}}{u^{2}+2} \cdot{\theta_{\rm E}}.
\end{equation} 
and second due to a likely blending with a luminous lens
\citep{1995A&A...294..287H,1995AJ....110.1427M,1995ApJ...453...37W}: 
 \begin{equation}
 \label{EQ:shiftlum}
\delta\boldsymbol{\theta_{c,\,lum}} = \frac{\boldsymbol{u}\cdot\theta_{\rm E}}{1+f_{\rm LS}}\,\frac{1+f_{\rm LS}(u^{2}+3-u\sqrt{u^{2}+4})}{u^{2}+2+f_{\rm LS}u\sqrt{u^{2}+4}}
.\end{equation}
This also changes the reference point to the center of light of the lens and the unlensed source.

For large impact parameters  (\(u\gg\sqrt{2}\)),  this simplifies to \citep{2000ApJ...534..213D}:
\begin{equation}
\label{EQ:shiftlum_approx}
        \delta\boldsymbol{\theta_{c,\,lum}} \simeq \frac{\delta\boldsymbol{\theta_{c}}}{1+f_{ls}} 
.\end{equation}
Please note that \(\boldsymbol{\delta\theta_{c}}\) and \(\boldsymbol{\delta\theta_{c,\,lum}}\) have different reference points, so only the amount of the positional shifts are connected via equation \ref{EQ:shiftlum_approx}.

For \(u\gg5\), the center of light is dominated by the major image. Therefore, the shift of the major images can be approximated by the shift of the center of light: 

 \begin{equation}
\delta\theta_{+} \simeq \delta\theta_{c} \simeq \frac{\theta_{\rm E}}{u}=\frac{\theta_{\rm E}^{2}}{\theta}\propto \frac{M_{\rm L}(\varpi_{\rm L}-\varpi_{\rm S})}{\theta},
 \label{Eq:shift_approx}
\end{equation}
where \(\theta = \lvert \boldsymbol{\theta} \rvert \).
This direct dependency on the mass shows the strength of astrometric microlensing. Further, due to the weaker dependence on the impact parameter, which is \(1/u\) rather than \(1/u^4\) for photometric microlensing, astrometric microlensing can be observed at larger impact parameters, which also results in longer time scales
\citep{1996AcA....46..291P,1996ApJ...470L.113M}. 
It can be described by \citep{2001PASJ...53..233H}  
\begin{equation} 
t_{aml} = 2\cdot t_{E}  \sqrt{\left(\frac{\theta_{E}}{\theta_{min}}\right)^{2} - u_{min}^{2}}
,\end{equation}
where \(\theta_{min}\) is the astrometric precision threshold of the used instrument. This assumes a rectilinear lens-source trajectory, and \(\theta_{E}/\theta_{min} \gg 5\), that means we can observe the effect in a regime where Equation~(\ref{Eq:shift_approx}) is valid.
With high-precision instruments this can be in the order of \(\theta_{min}= 0.1\, \mathrm{mas}\), which may lead to time-scales of many months or even a few years.  
Further, the weaker $u$ dependence also allows more confident predictions of astrometric-microlensing events.

\begin{figure}
    \centering
    \includegraphics[width=0.95\linewidth]{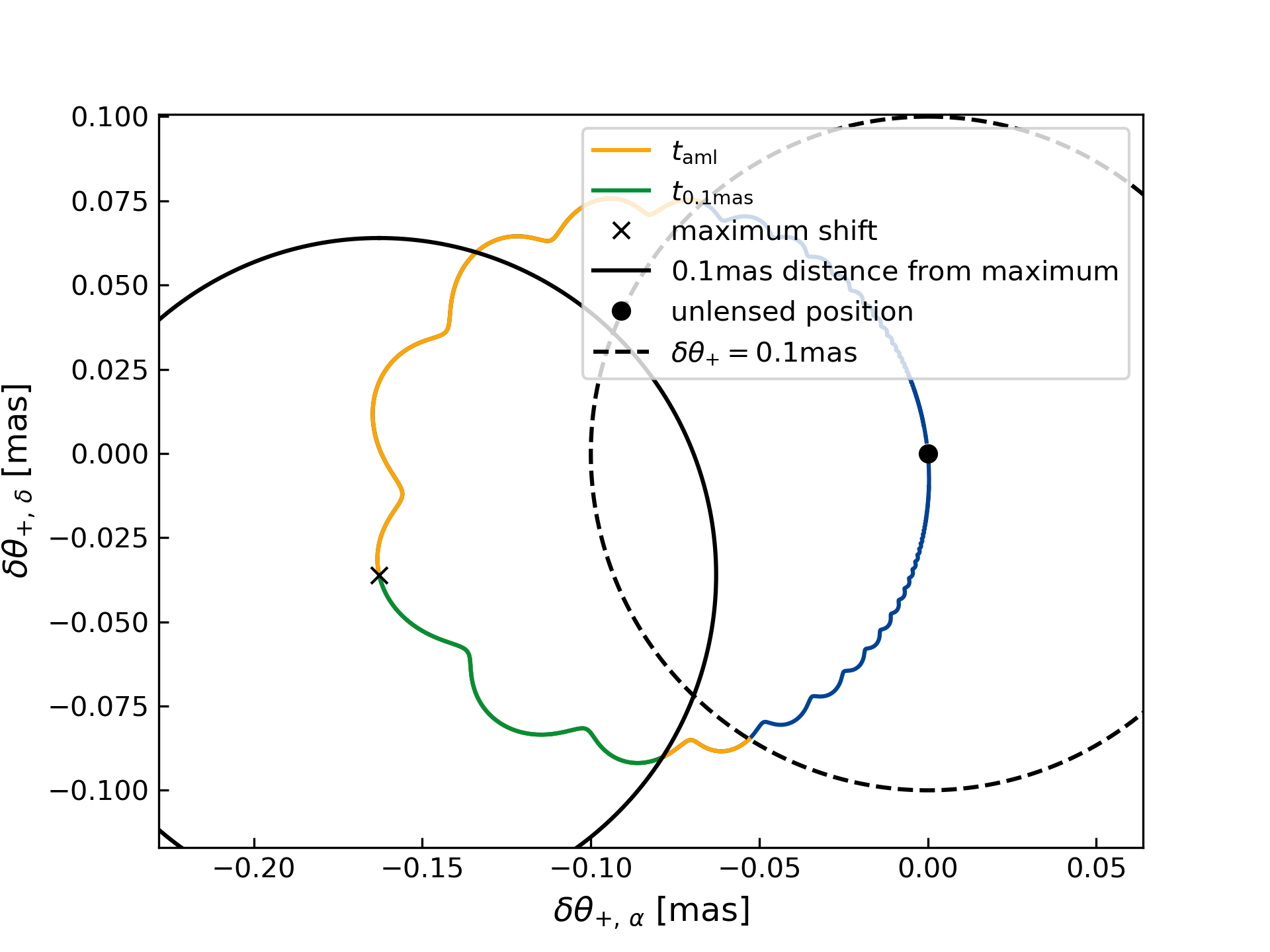}
    \caption{Geometric illustration for the different time scales. The wobbling line shows the position of the major image with respect to the unlensed position of the source at the same epoch. 
    During the event the the major images moves along this line in a clockwise or anticlockwise direction (clockwise for the shown event). 
    The wobbling motion is caused by the parallactic motion of the lens.
    The black dashed and solid circle are centered on the origin, and the position of the major image at the epoch of the closest approach. 
    Both have an radius of \(0.1\mas\).  
    \(t_{aml}\) describes the duration where the major image is outside of the dashed circle (orange + green, part of the line). 
    \(t_{0.1\mas}\) describes duration between entering the solid circle an reaching the center (green part of the line)  
    Similar to \(t_{0.1\mas}\), \(t_{50\%}\) descries how long takes if circle would have a radius of \(0.5\delta\theta_{+}\).}
    \label{Fig:time_scales}
\end{figure}

On the other hand, the long time scales can lead to small variation within a given time, which can be hard to measure. To indicate how fast the position changes due to microlensing, we estimate two time scales \(t_{0.1\mas}\) and \(t_{50\%}\).
These indicates how long it takes until \(\delta\boldsymbol{\theta}\) change by \(0.1\mas\) (i.e.  \(\lvert\delta\boldsymbol{\theta}_{+}(T_{CA})-\delta\boldsymbol{\theta}_{+}(T_{CA}-t_{0.1\mas})\rvert = 0.1\mas\), where \(T_{CA}\) is the epoch of the closets approach) and \(0.5\cdot \,\delta\theta_{+}\) from the maximum deflection, respectively. 
We note that  \(t_{50\%}\) is longer than  \(t_{0.1\mas}\) for events with an shift larger than \(0.2\mas\) and vice versa if the maximum shifts is smaller than \(0.2\mas\). 
Figure \ref{Fig:time_scales} shows the positional shift due to astrometric microlensing.
\(t_{aml}\) corresponds the duration outside of the dashed circle 
(\(\delta\theta_{+} > 0.1\mas\), orange and green part) 
and \(t_{0.1\mas}\) corresponds to the duration between entering the solid circle 
(\(0.1\mas\) around the point of the maximum positional shift) 
and reaching the maximum positional shift (green part). 
We estimates those time scales numerically using a one-week grid.
If we encounter a duration below 2 weeks, we switch to a 1 day grid. Later, typically effects events with very high proper motions compared to their impact parameters squared (\(\mu_{rel}/u_{min}^{2} > \sim 5\mpy\)).

\section[Prediction of microlensing events]{Prediction of microlensing events\footnote{The source code for this analysis is publicly available at \\ \urlpython.}}
\label{Section:Search}

\begin{figure*}
\center
	\includegraphics[width=0.9\linewidth]{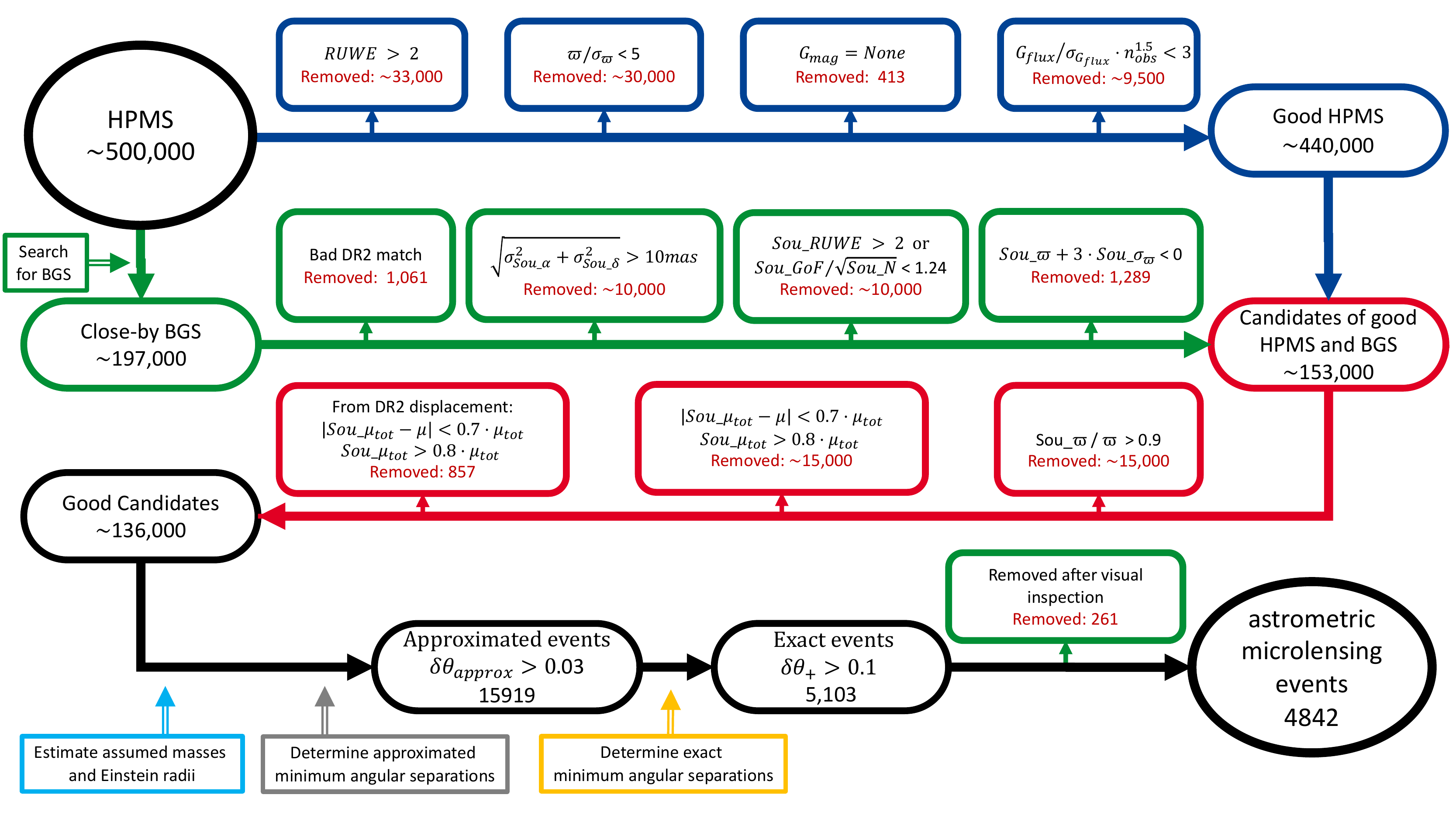}
\caption{Illustration of the selection process. The search for astrometric-microlensing events is divided into four major parts. The selection of good high proper-motion stars (blue), the selection of good background stars (green), the exclusion of co-moving stars (red), and the determination of the closest approach and estimation of the expected microlensing effect (black).
The search for close by BGS is performed on the sample of \(\sim 500,000\) HPMS and the selection of good HPMS and BGS is performed independently from each other. 
The illustration lists the different selection criteria, with the number of excluded HPMS, BGS, Candidates or events. In each process, stars can be excluded due to multiple criteria. The small rectangular boxes indicate where the major computation is performed}
\label{Fig:Method} 
\end{figure*}

In principle, our search follows the method we described in \Kt or in \cite{2011A&A...536A..50P}.
However, we made major adjustments in the selection of potential BGS and high-proper-motion stars (HPMS). The selection processed is shown in Fig~\ref{Fig:Method}.

\subsection{List of high-proper-motion stars}
\label{subSection:HPMS}

\begin{figure}
\center
\includegraphics[width=0.95\linewidth]{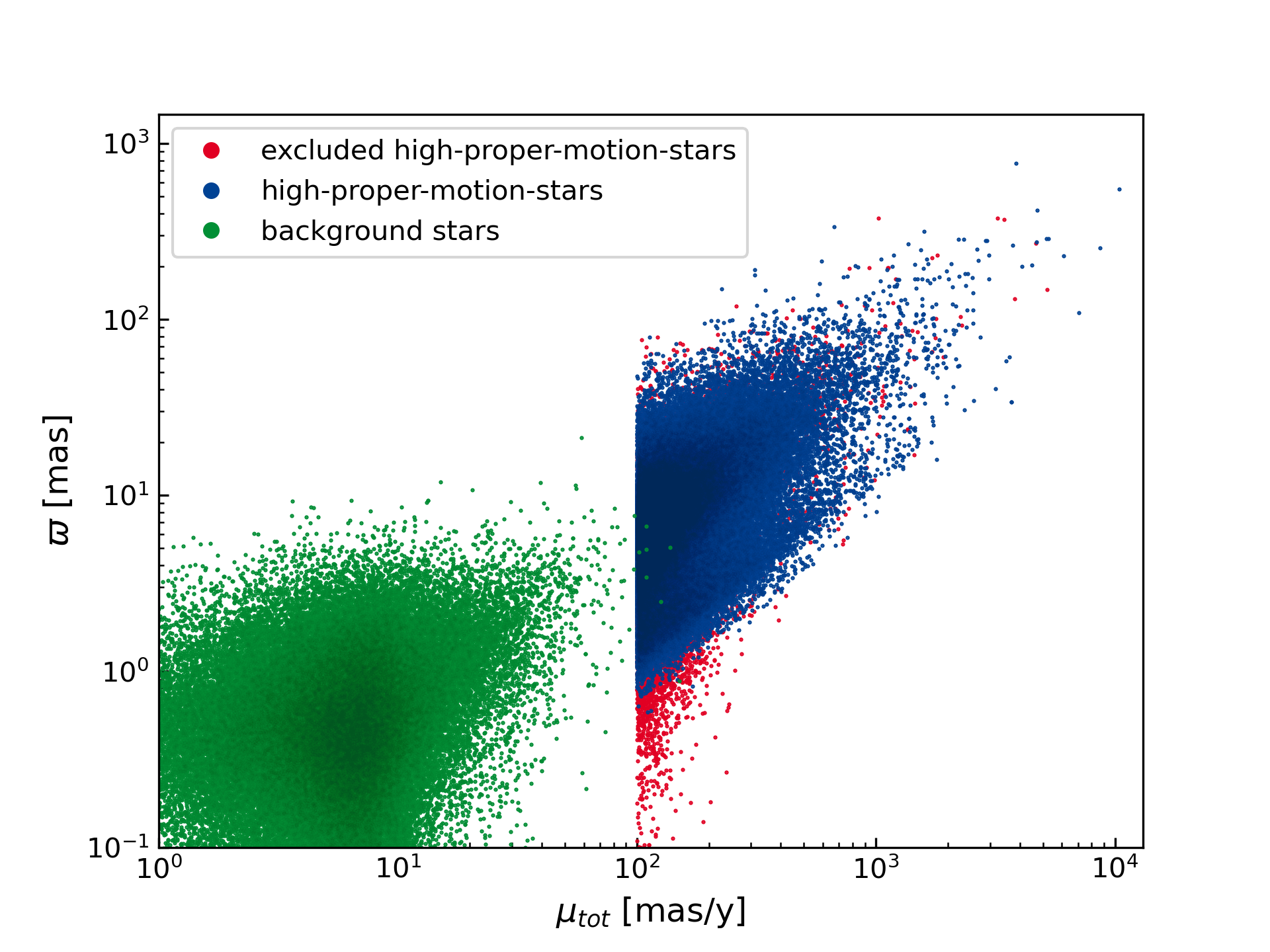}
\caption{Proper motions (\(\mu_{tot}\)) and parallaxes (\(\varpi\)) for all used HPMS(Blue), excluded HPMS(red), and BGS (green). The high number of HPMS excluded with small parallaxes, are due to the selection of HPMS with \(\varpi/\sigma_{\varpi}>5\).
The BGS and HPMS shows a clear separation, the few BGS \(\mu_{tot}>100\mpy\) are passed by HPMS with even higher proper motions.}
\label{Fig:HPMS} 
\end{figure}

\begin{figure}
\center
\includegraphics[width=0.95\linewidth]{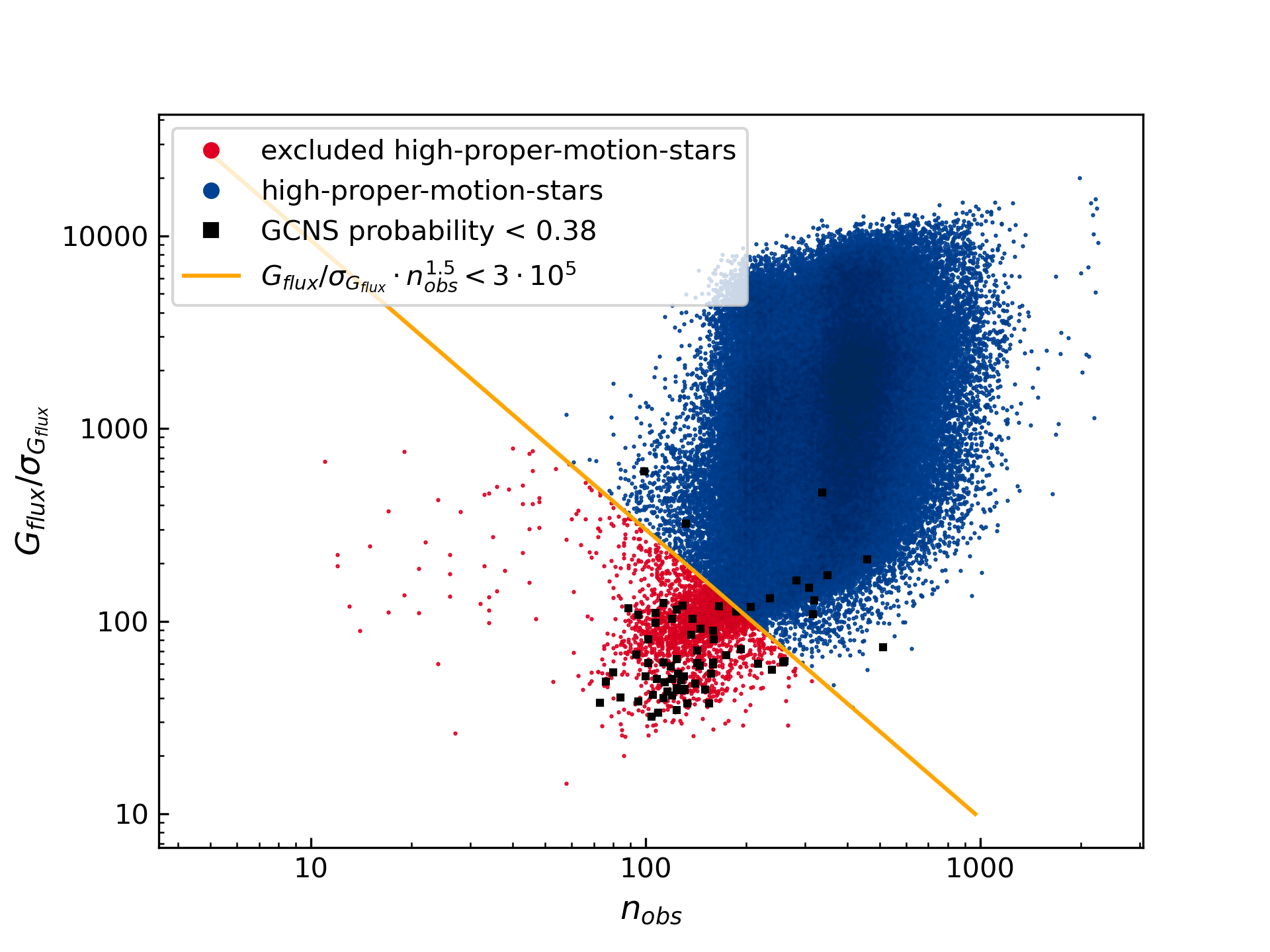}
\caption{Number of photometric observations by \Gaia (\(n_{\rm obs}\)) and significance of the G flux (\(G_{\rm flux}/\sigma_{G_{\rm flux}}\)) for all potential high-proper-motion stars.  The black squares indicate the HPMS rejected in the GCNS. The HPMS with low fidelity values also cluster in the bottom left region. The excluded HPMS are plotted in red, and the HPMS which passes this criterion are plotted in blue.}
\label{Fig:nVSsig} 
\end{figure}

We started with the selection of HPMS as potential lenses. Compared with our previous study, we extended our search to HPMS  with absolute proper motions  \(\mu_{tot} = \sqrt{\mu_{\alpha^{*}}^{2}+\mu_{\delta}^{2}} > 100 \mpy\).
The limit of \(100\mpy\) still allows a clear separation between background stars and co-moving stars (see Fig.~\ref{Fig:HPMS}).  In
\eDR, about \(500,000\) sources fulfill this criterion.
As mentioned by \cite{refId10} and \cite{refId7}, the \eDR contains a small fraction of erroneous data. For example, about 1000 HPMS show a significantly negative parallax. 
To exclude such sources, and to ensure a good astrometric solution in \eDR we applied the following quality cuts. We first excluded all HPMS with a re-normalized unit weight error (\texttt{ruwe}) above~2. This deleted about \(33,000\) HPMS. Further, we excluded HPMS with the significance of the parallax \((\varpi/\sigma_{\varpi})\) less than 5. About \(30,000\) do not pass this criterion. While the \texttt{ruwe} limit excludes mostly bright HPMS (\(G< 18 \MAG\)), faint HPMS (\(G>18 \MAG\)) are mostly excluded by the parallax criterion. Finally, we excluded 413 sources, since \eDR does not provide a G magnitude for them.

In order to validate our selection, we cross-matched our list of HPMS with the \Gaia Catalogue of Nearby Stars \citep[GCNS,][]{refId1}. We found \({\sim}144,000\) common stars in the GCNS proper and \({\sim} 50,000\) common stars in the GCNS rejected catalogue. The latter are almost exclusively rejected due to parallaxes between  \(8 \mas\) and \(10\mas\). Only 95 common sources are rejected due to a low GCNS probability. 

In \Kt we also found that suspicious data cluster in the number of photometric observations \((n_{obs})\) - 
G-flux significance \((G_{flux}/\sigma_{G_{flux}})\) space. This is also true for most of the 95 HPMS rejected by the GCNS (see Fig.~\ref{Fig:nVSsig}).
Consequently, we excluded \({\sim} 9,500\) HPMS with
\begin{equation} 
G_{flux}/\sigma_{G_{flux}} \cdot n_{obs}^{1.5}< 3 \cdot 10^{5}.
\end{equation} 
This limit is indicated by the line in Fig.~\ref{Fig:nVSsig}.

Using \DRtwo, in \Kt we had further found two populations of erroneous data with typical tangential velocities of \((v_{tan} \simeq 4\,\mathrm{km/s})\) and \((v_{tan} \simeq 11\,\mathrm{km/s})\), respectively.
In \eDR, we could not detect these, nor any other unexpected population  (see Fig.~\ref{Fig:HPMS}). Consequently, we did not apply a filter on the tangential velocity. About \(440,000\) HPMS pass all 5 criteria. For these, we searched for background sources close to their expected paths on the sky.

\cite{2021arXiv210111641R} determined a classifier in order to indicate sources with spurious astrometric data in \eDR. Using a machine-learning algorithm, they determined a fidelity value between 0 and 1 for each source with a 5-parameter solution in \eDR. A value of 1.0 means a perfectly trustable solution and the lowest possible value of 0.0 indicates a lot of issues in the astrometric solution. Further, they propose a cut at 0.5 to differentiate between ``good'' and spurious sources. Only 126 of all HPMS which passed our five criteria have a fidelity value below 0.75 and none of those result in a predicted event, even without any cuts on the fidelity value. On the other hand, a large fraction of the excluded HPMS (\(97\%\)) have a fidelity value above 0.75. However, to ensure a good prediction, we kept the above-described criteria.

\subsection{Potential background stars}
\label{subSection:BGS} 
\begin{figure}
\center
\includegraphics[width=0.95\linewidth]{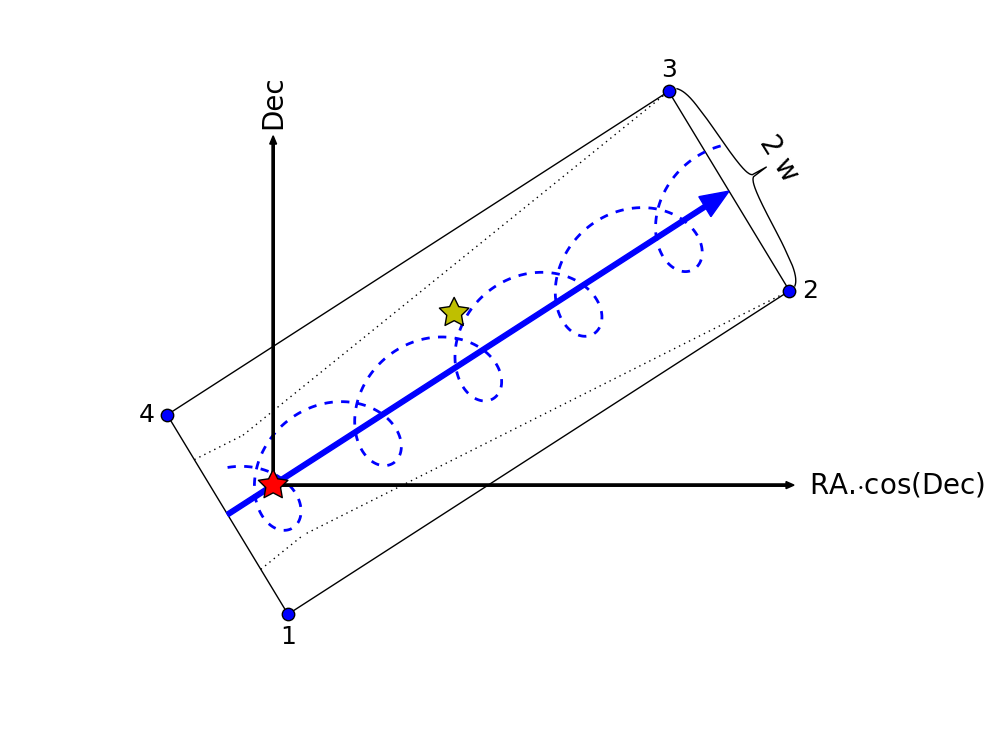}
\caption{Illustration of the rectangular box used in our search for background stars. The thick solid blue line indicates the direction of the lens proper motion  (red star), and the origin is set to the J2016.0 position of the lens. The blue dashed line indicates the real motion, which includes the parallax (only five years are shown, and not to scale). The real size is defined by the position of the lens in J2010.5 and J2066.0 and a half width of \(w = 7''\). When the J2016.0 position of a background star (yellow star) is within our rectangle, it is considered as candidate.  To account for the proper motion of the background source, the widening shape (dotted black line) would be more physically accurate. Plot taken from \Kt.}
\label{Fig:Window} 
\end{figure}

\begin{figure}
\center
\includegraphics[width=0.95\linewidth]{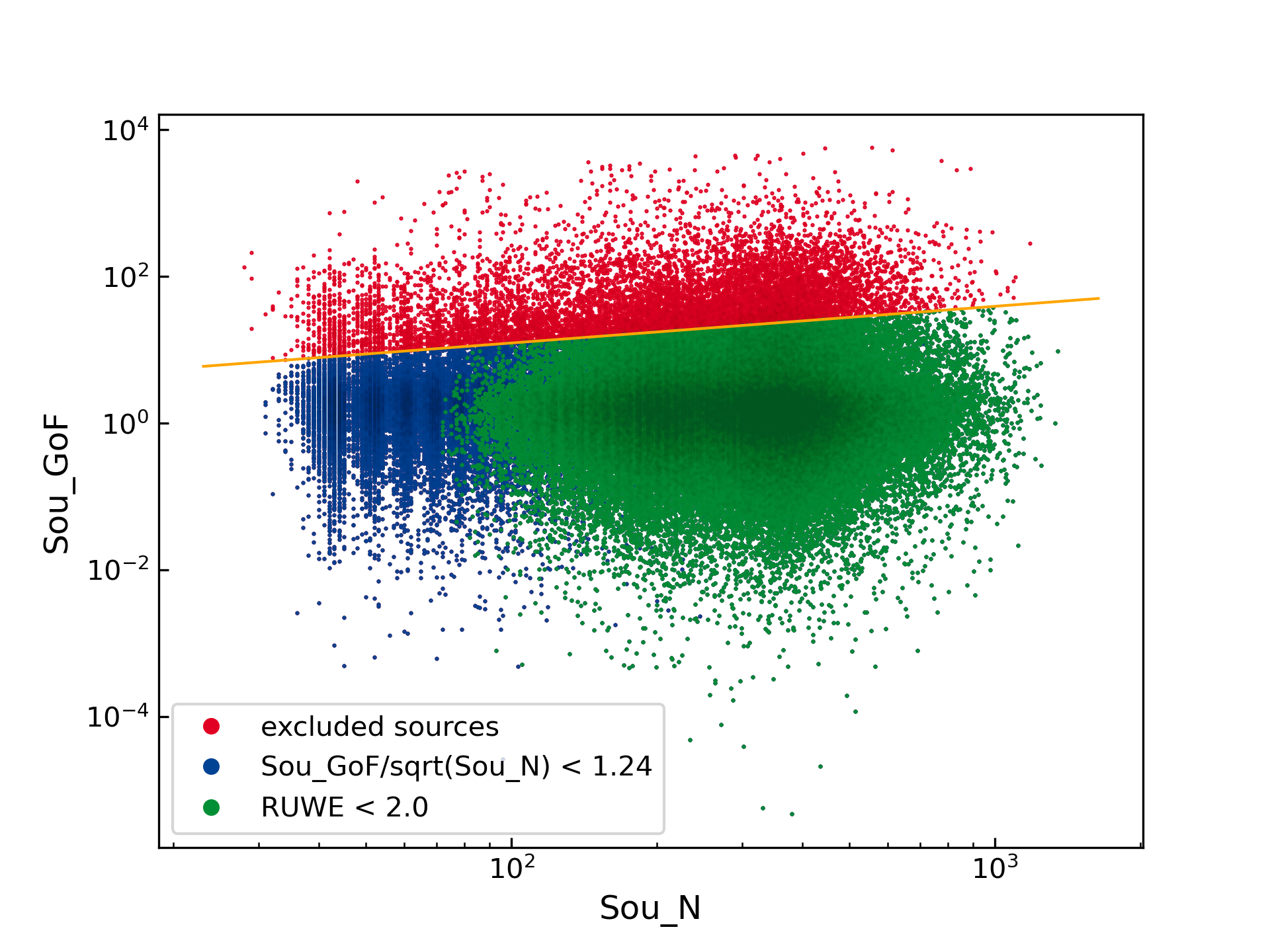}
\caption{Goodness of fit for \Gaia's astrometry, versus number of good astrometric observations. BGS with a \texttt{ruwe} less than 2 are plotted in green. These shows a sharp edge which we used to determine a limit for stars which do not have a 5-parameter solution. Such sources which pass this criterion are shown in blue, and the excluded background sources are plotted in red.}
\label{Fig:GoF} 
\end{figure}

As in \Kt, for each HPMS we defined a rectangular box on the sky, using the J2010.5 and J2066.0 positions and a half-width of \(w= 7\mas\) (see Fig.~\ref{Fig:Window}) to search for suitable background sources. In the following the pair of a foreground HPMS and background star (BGS) is called a ``candidate'', where the BGS parameters are labelled with the prefix ``\(\Sou\)''.

Among all stars in the boxes we first excluded BGS with significantly negative parallaxes  (\(\Sou\varpi+ 3\cdot \Sou\sigma_{\varpi} < 0\mas\)),
	$G$ magnitudes fainter than \(G = 21.5\) and without G magnitudes in \eDR.  
	Second, we excluded all BGS with a \texttt{ruwe} above~2. This criterion can only be applied for BGS with a five-parameter solution in \eDR. However, the \texttt{ruwe} strongly correlates with the astrometric goodness-of-fit over the square root of the number of good astrometric microlensing along-scan observations (\(\Sou GoF/\sqrt{\Sou N}\), see Fig.~\ref{Fig:GoF}). 
Hence, if a five-parameter solution does not exists in  \eDR, we only considered sources with \(\Sou GoF/\sqrt(\Sou N)\) < 1.24, which is equivalent to \({\rm \texttt{ruwe}} < 2\).

\begin{figure}
\center
\includegraphics[width=0.95\linewidth]{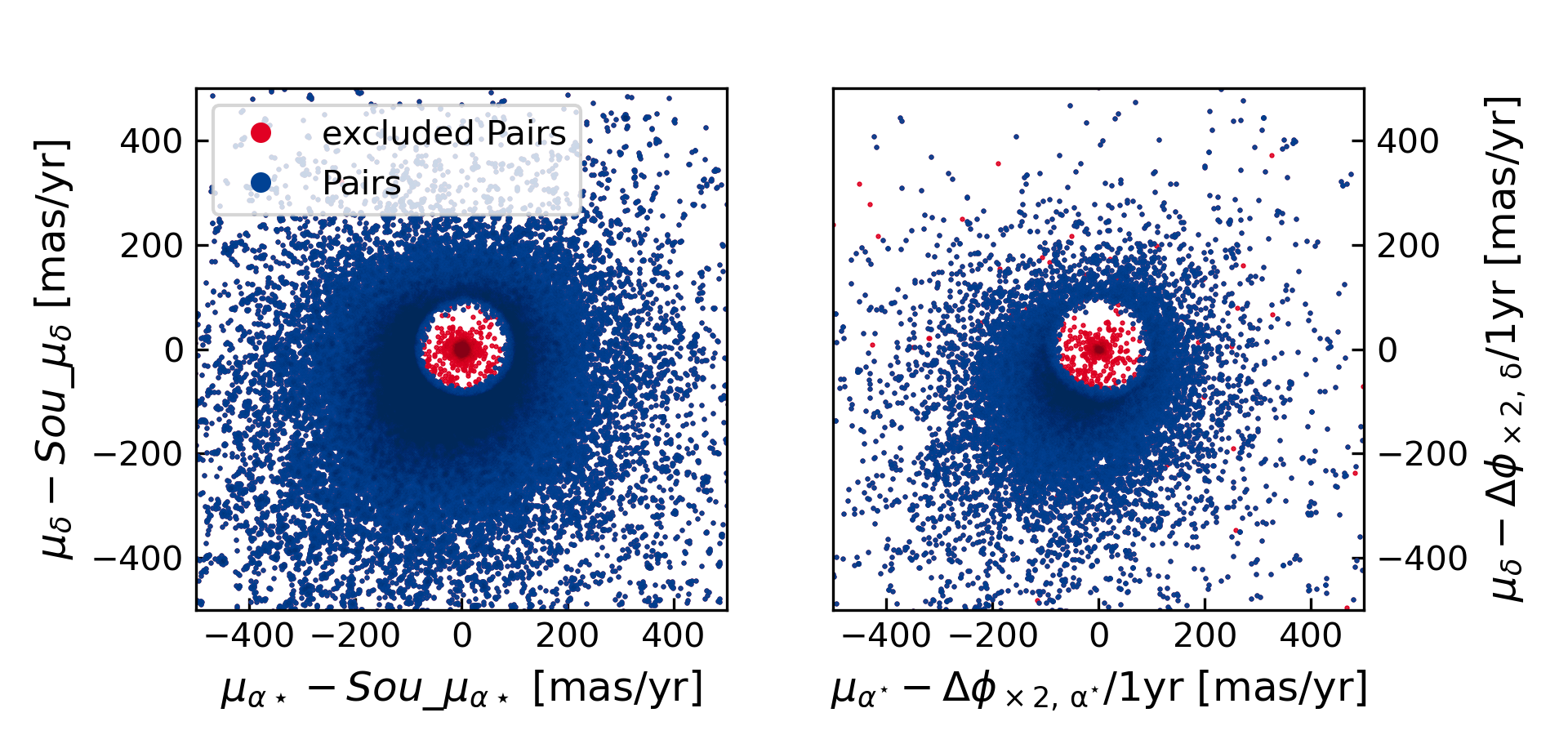}
\caption{Difference in the proper motion  between HPMS and BGS. The left panel shows the stars with a proper motion listed in \eDR, while in the right panel we used the positional displacement between DR2 and DR3 as an estimate for the proper motion.
The excluded candidates are plotted in red.
The sparsely scattered candidates with large difference in the right panel are excluded since the absolute value of the proper motions are similar. 
}
\label{Fig:simPM} 
\end{figure}
	
In order to avoid physical binary stars as well as co-moving pairs of stars, we apply various filters on the parallax and proper motion. We first consider only candidates with the parallax of the source smaller than 0.9 times the parallax of the lens. i.e: 
\begin{equation}
	 (\Sou \varpi/\varpi < 0.9)
	 \label{Eq:simlimit1}
\end{equation}	
	The main purpose of this cut is to avoid imaginary Einstein radii (i.e. the "source star" being closer than the "lens star"), since the proper motion is more effective to distinguish between background and co-moving stars, especially for distant stars \((\varpi < 10 \mas)\).
	For the  difference in proper motion we applied the following two criteria: 
\begin{align} 
	\lvert \Sou \boldsymbol{\mu} - \boldsymbol{\mu} \rvert > 0.7 \cdot \mu_{tot}\\
	\Sou\mu_{tot}  < 0.8 \cdot \mu_{tot}
\label{Eq:simlimit2}
\end{align}
The difference in proper motion between HPMS and BGS are shown in the left panel of Fig.~\ref{Fig:simPM}
\subsubsection{Displacement between DR2 and eDR3 as proxy for missing proper motions}
\label{SubSubsection:DR2pm}

\begin{figure}
\center
\includegraphics[width=0.8\linewidth]{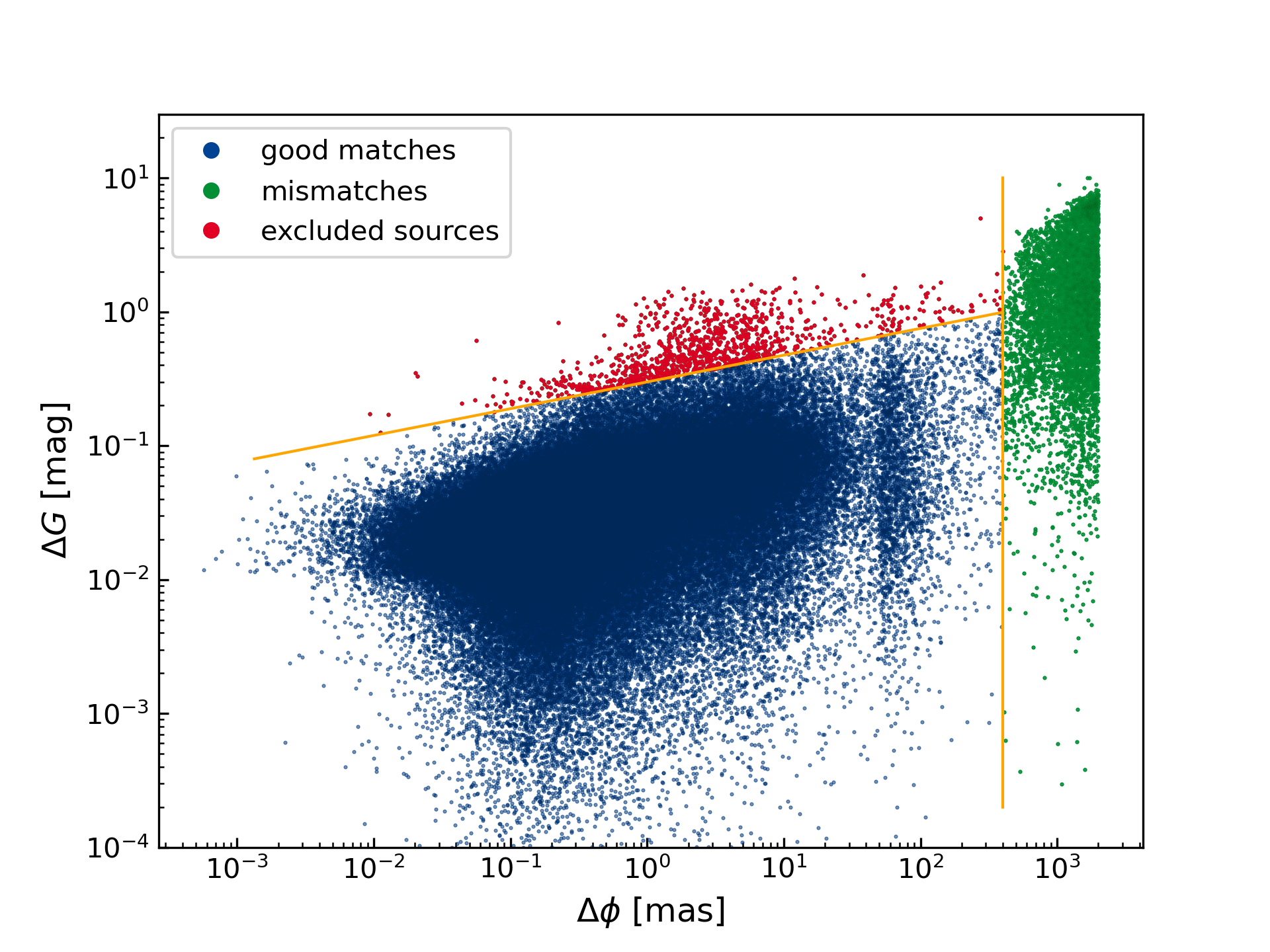}
\caption{Absolute value of the magnitude difference (\(\Delta G \)) versus angular separation (\(\Delta \phi \)) for the cross match between \eDR and DR2, as listed in the DR2 neighbourhood catalogue. The grey dots in the background show the distribution for a random sample.
 The green dots (\(\Delta\phi>400\mas\)) are interpreted as false matches, these are ignored, while the good matches are shown in blue. The population of stars with \(\Delta\phi\) around \(100\mas\)
are interpreted as co-moving stars, without a 5-parameter solution in \DRtwo. 
 The red dots, indicate sources excluded due to a mismatch in the magnitudes.
The orange lines indicate the limits for our classification. 
We note hat including more false matches lead to a stricter selection of candidates.
}
\label{Fig:DR2Match} 
\end{figure}

\begin{figure*}
\gridline{
    \fig{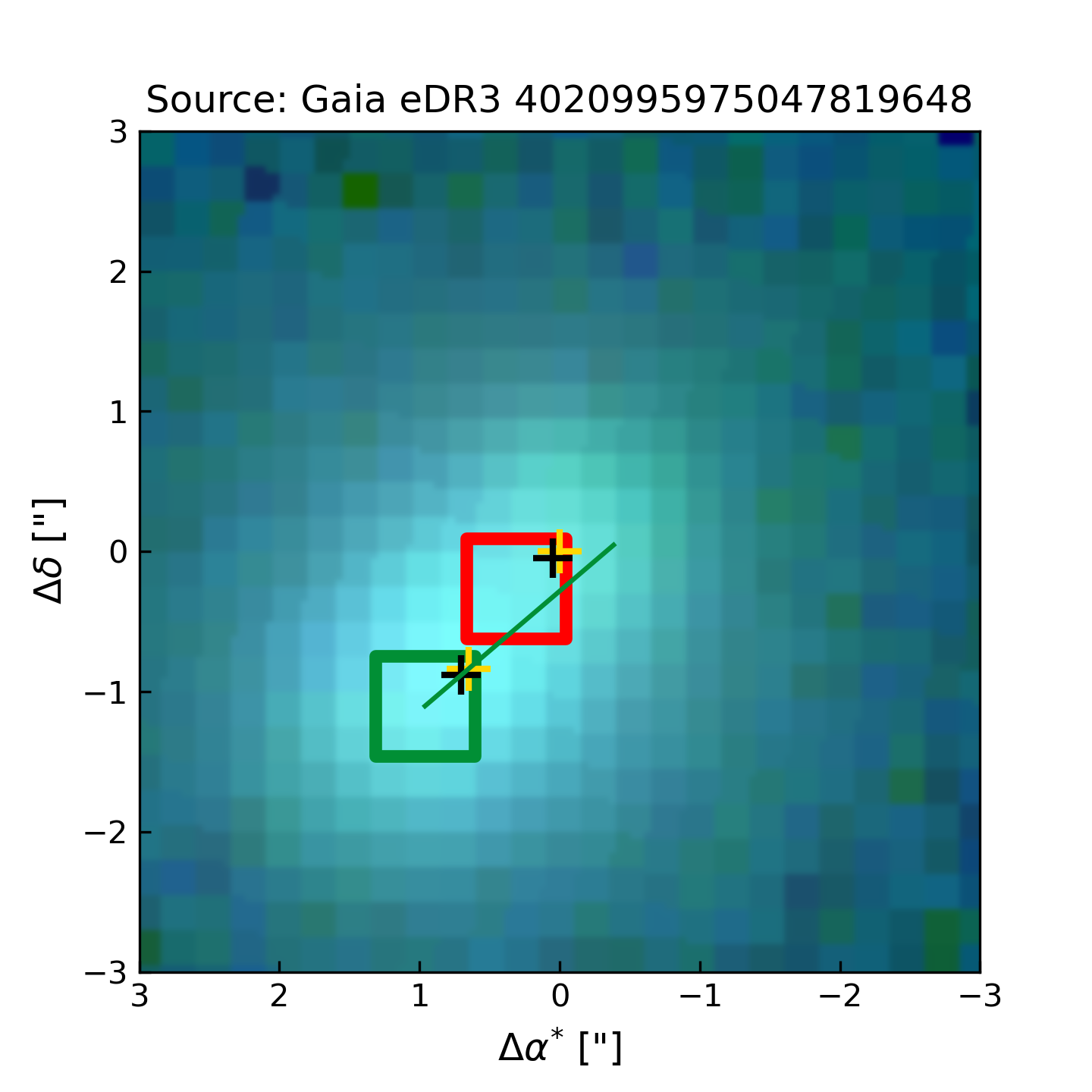}{0.24\textwidth}{(a) PanSTARRS}
    \fig{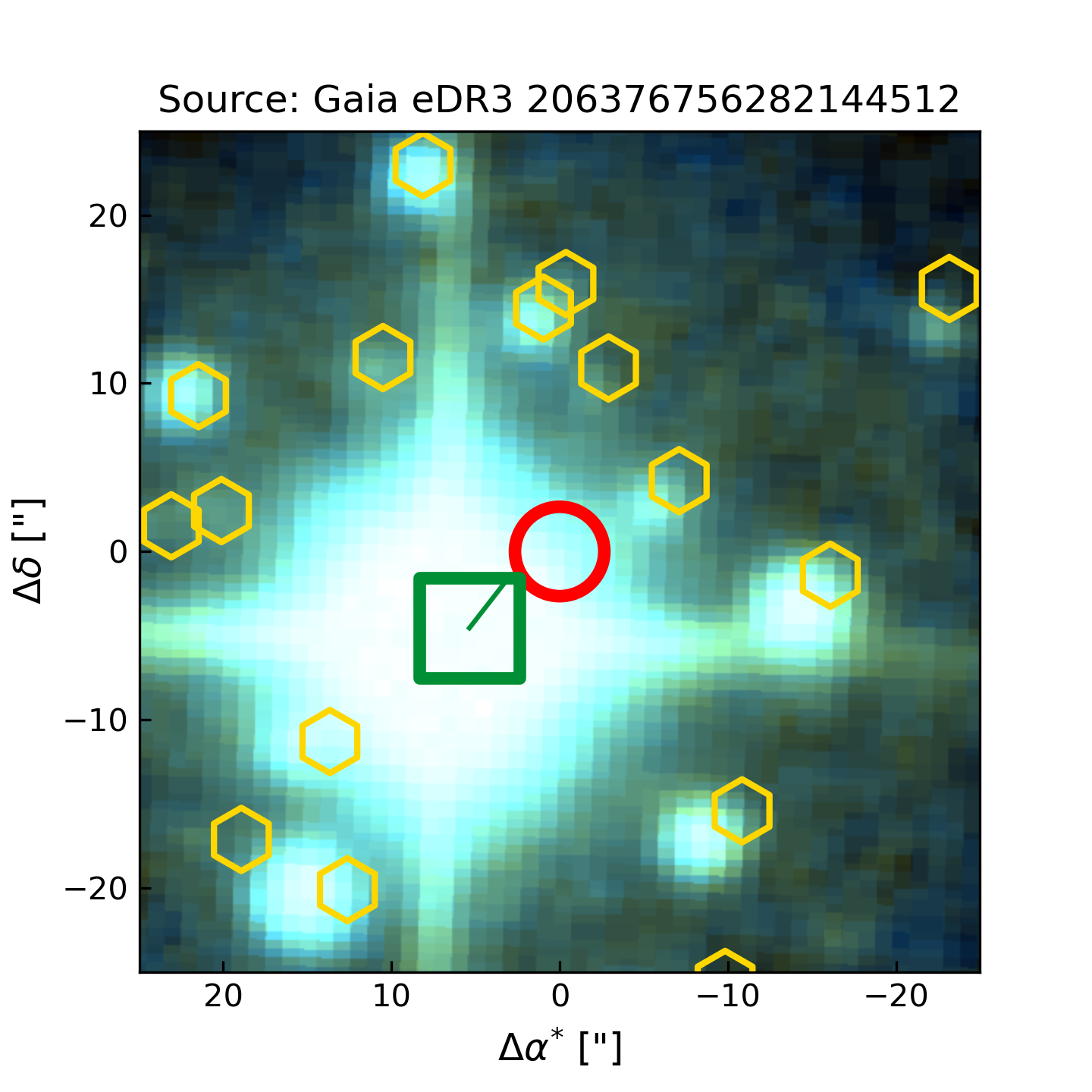}{0.24\textwidth}{(b) DSS RED-BLUE}
    \fig{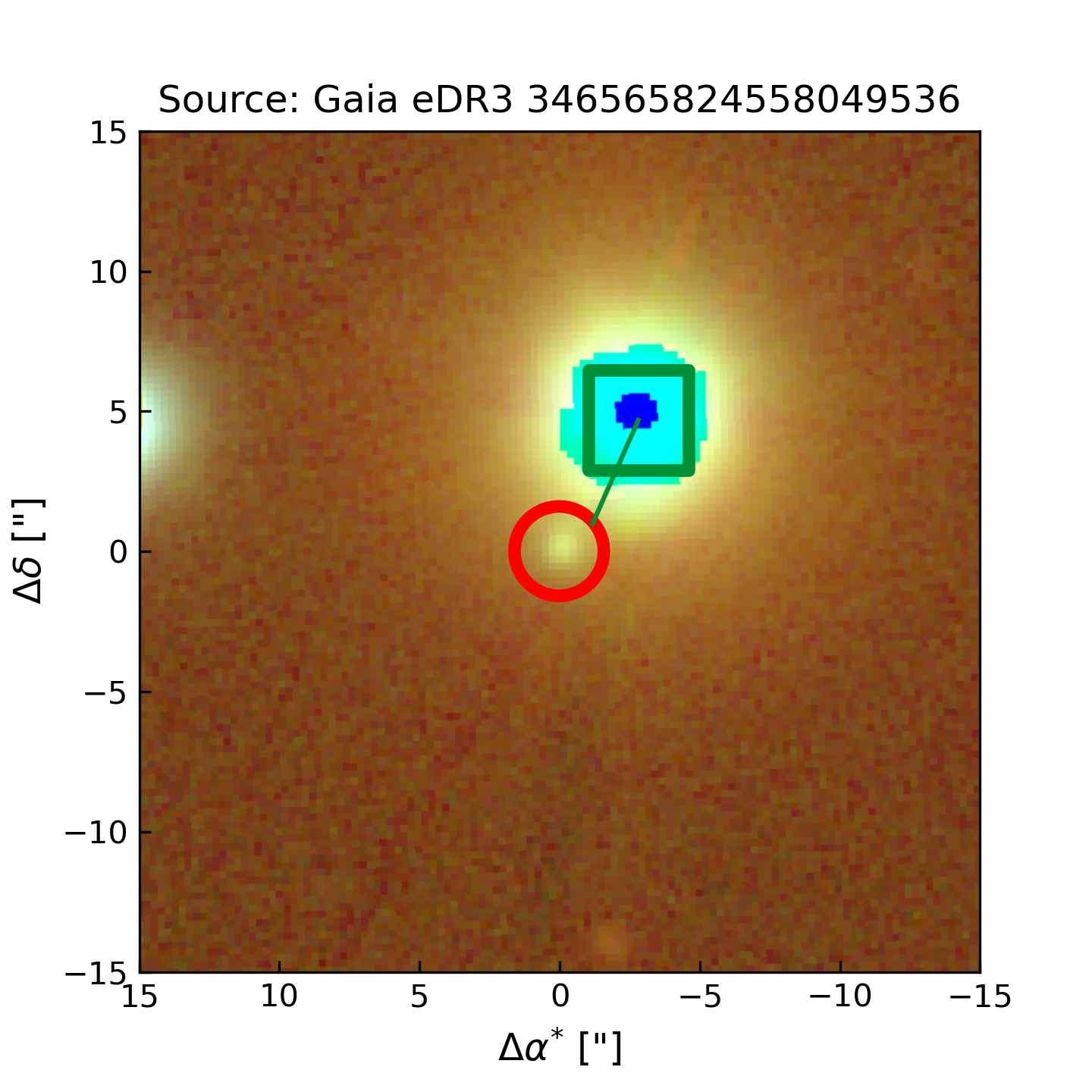}{0.24\textwidth}{(c) PanSTARRS i-r-g}
    \fig{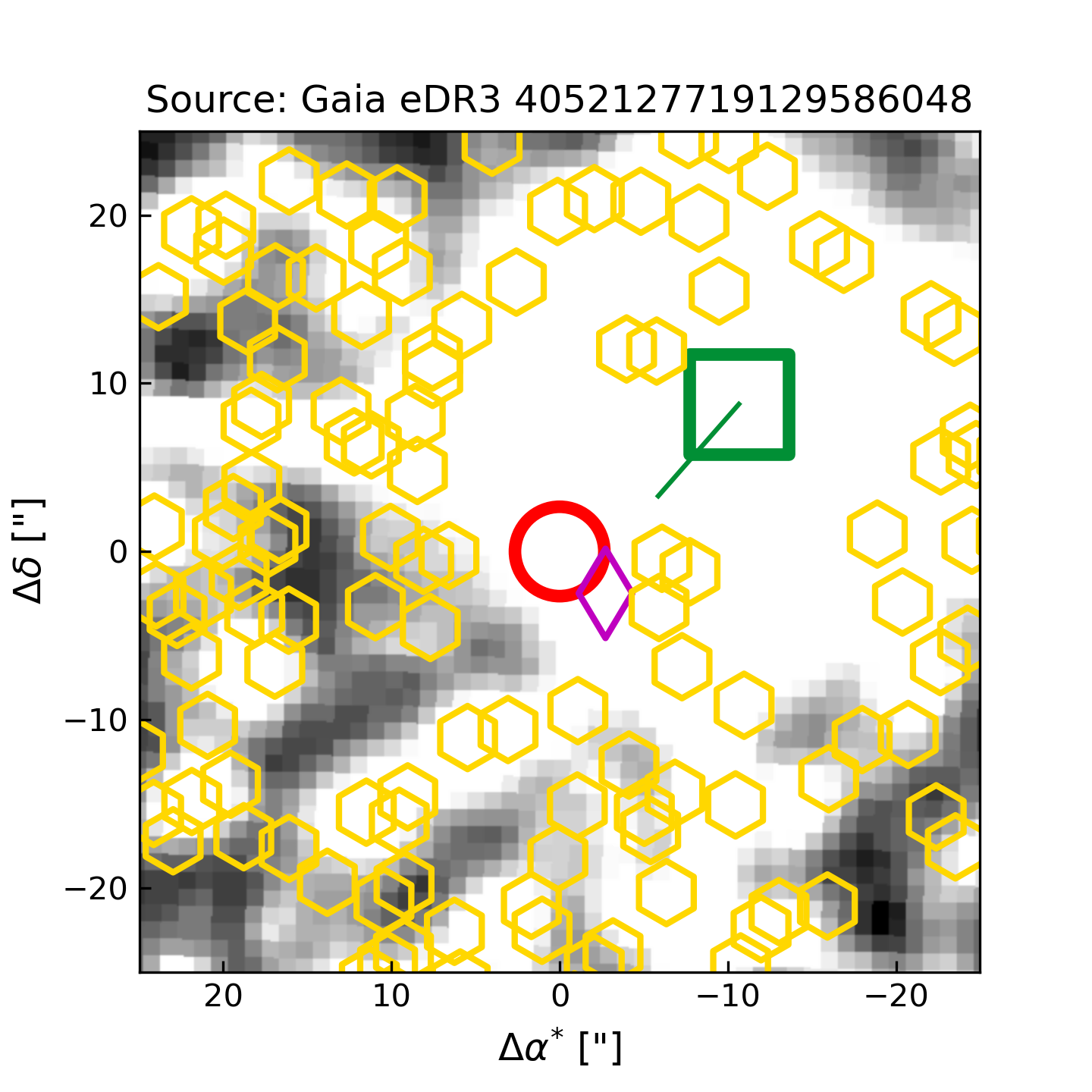}{0.24\textwidth}{(d) DSS RED}
}
\gridline{
    \fig{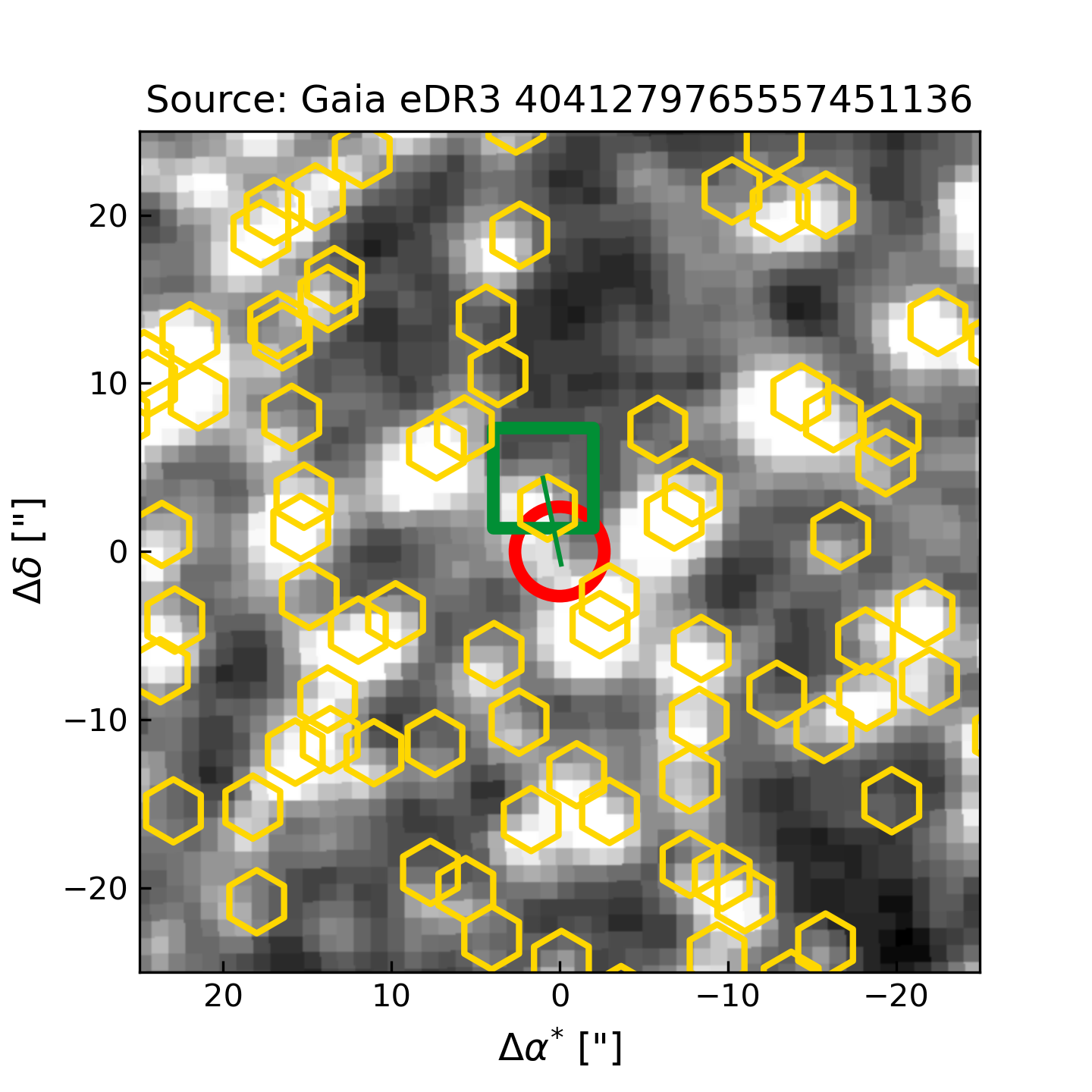}{0.24\textwidth}{(e) DSS RED}
    \fig{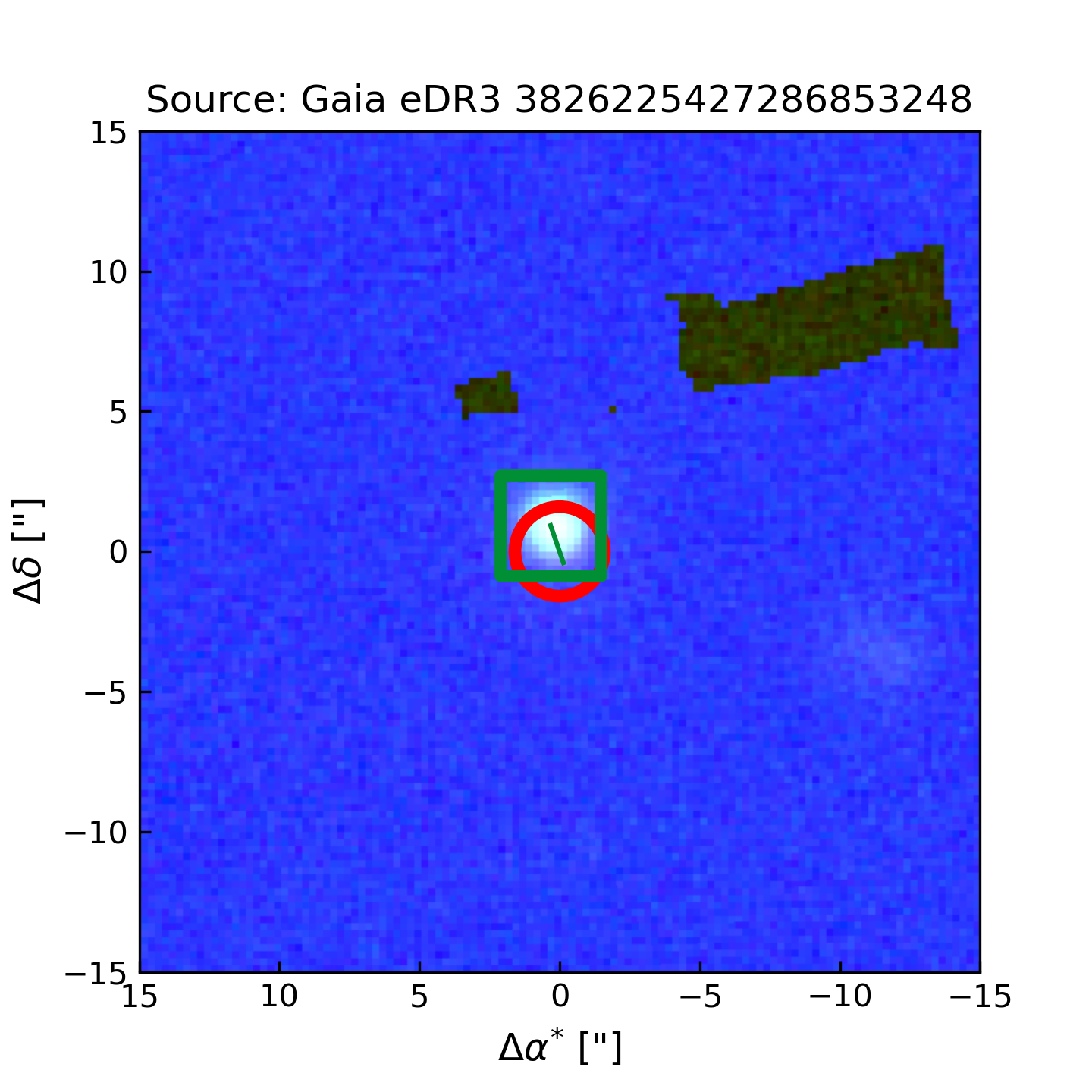}{0.24\textwidth}{(f) PanSTARRS i-r-g}
    \fig{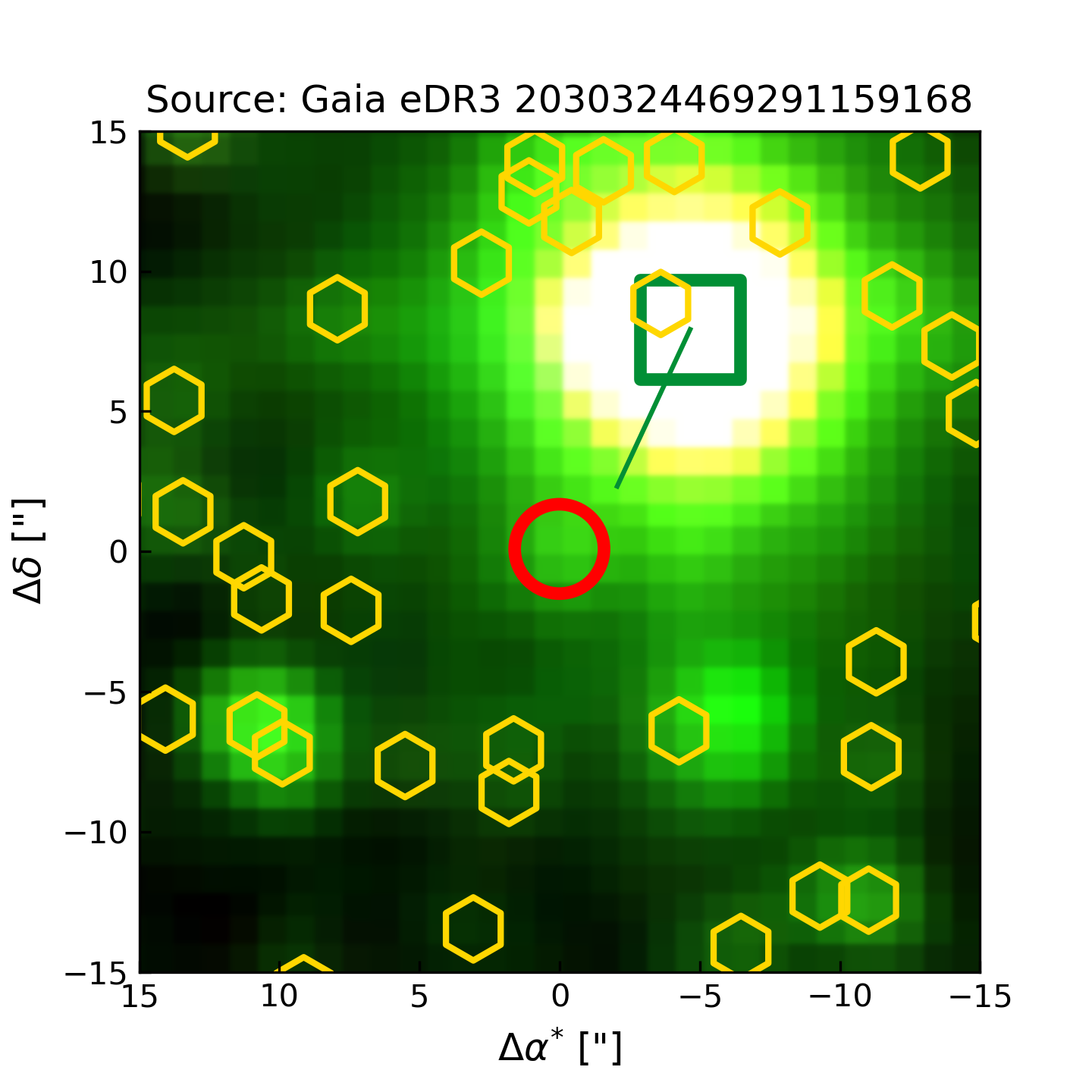}{0.24\textwidth}{(g) 2MASS Ks-H-J}
    \fig{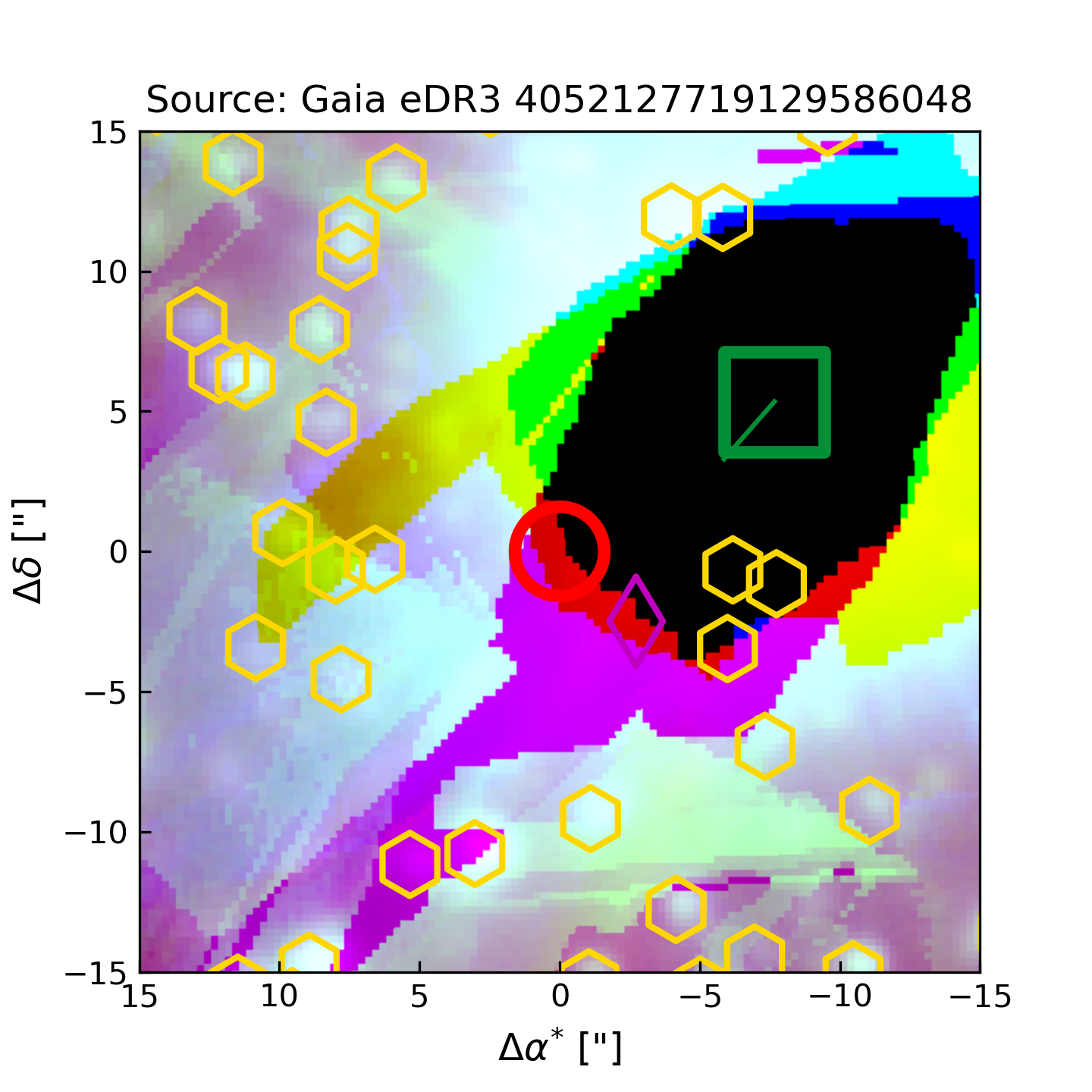}{0.24\textwidth}{(h) PanSTARRS i-r-g} %
    }
\gridline{
    \fig{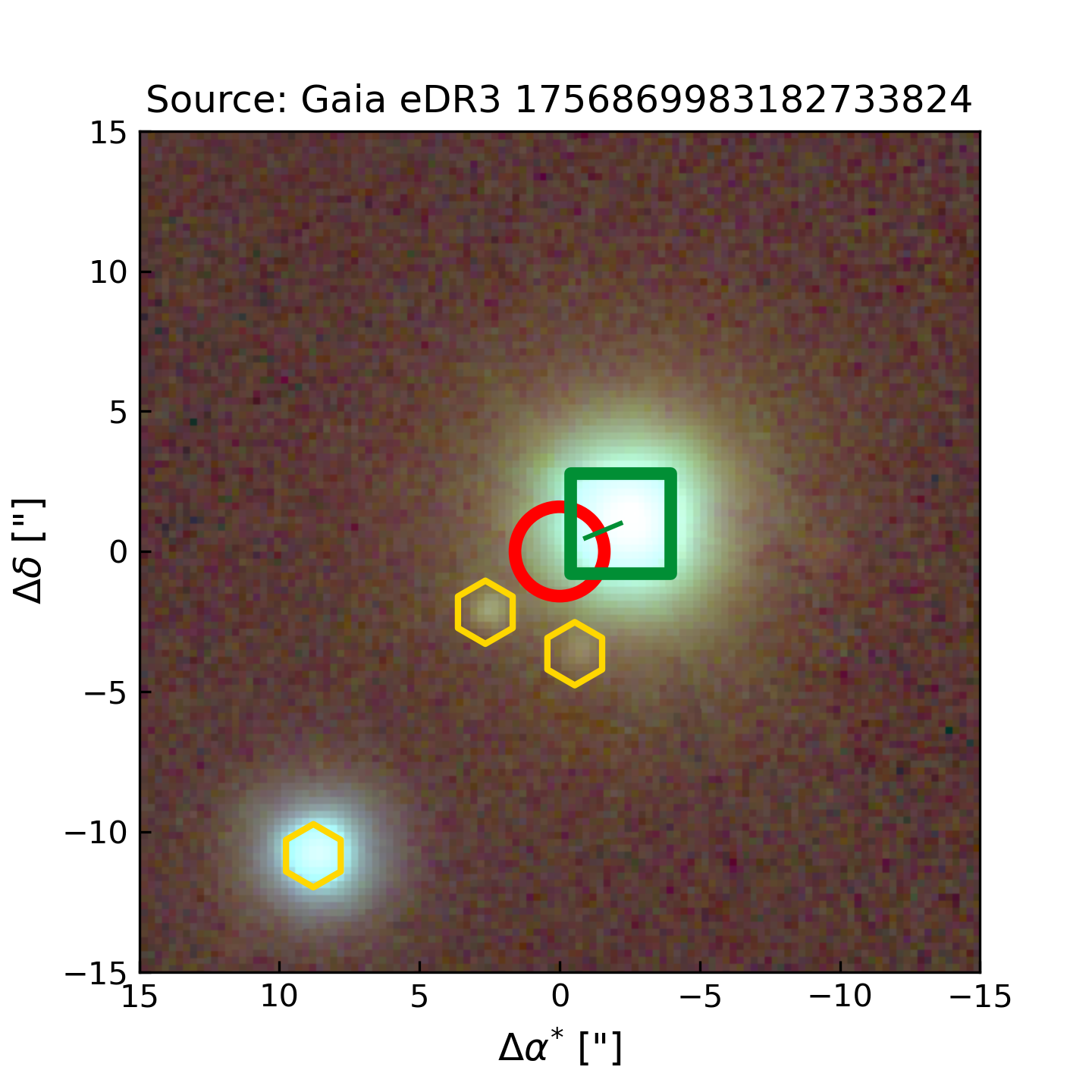}{0.24\textwidth}{(i) PanSTARRS i-r-g}
    \fig{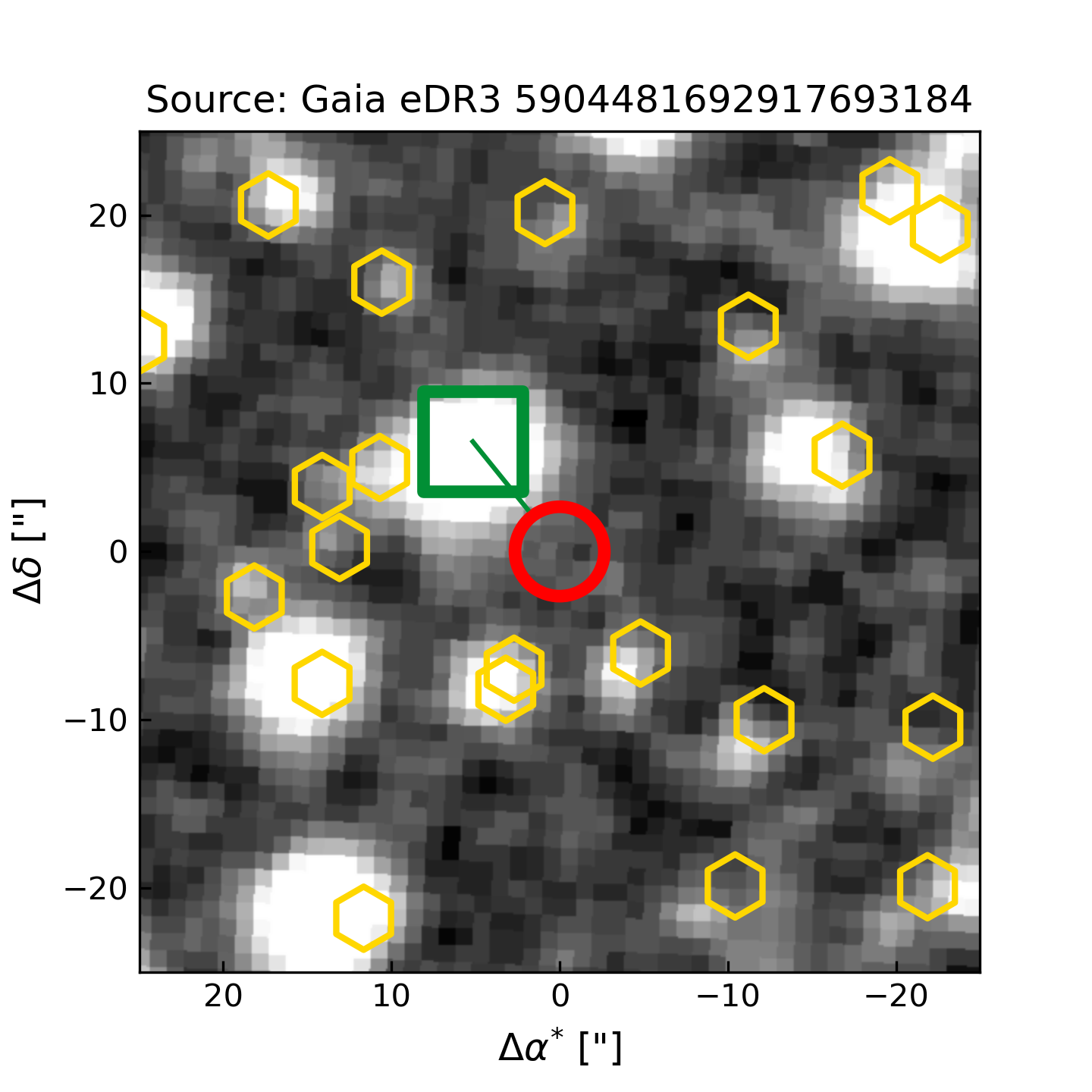}{0.24\textwidth}{(j) DSS RED}
    \fig{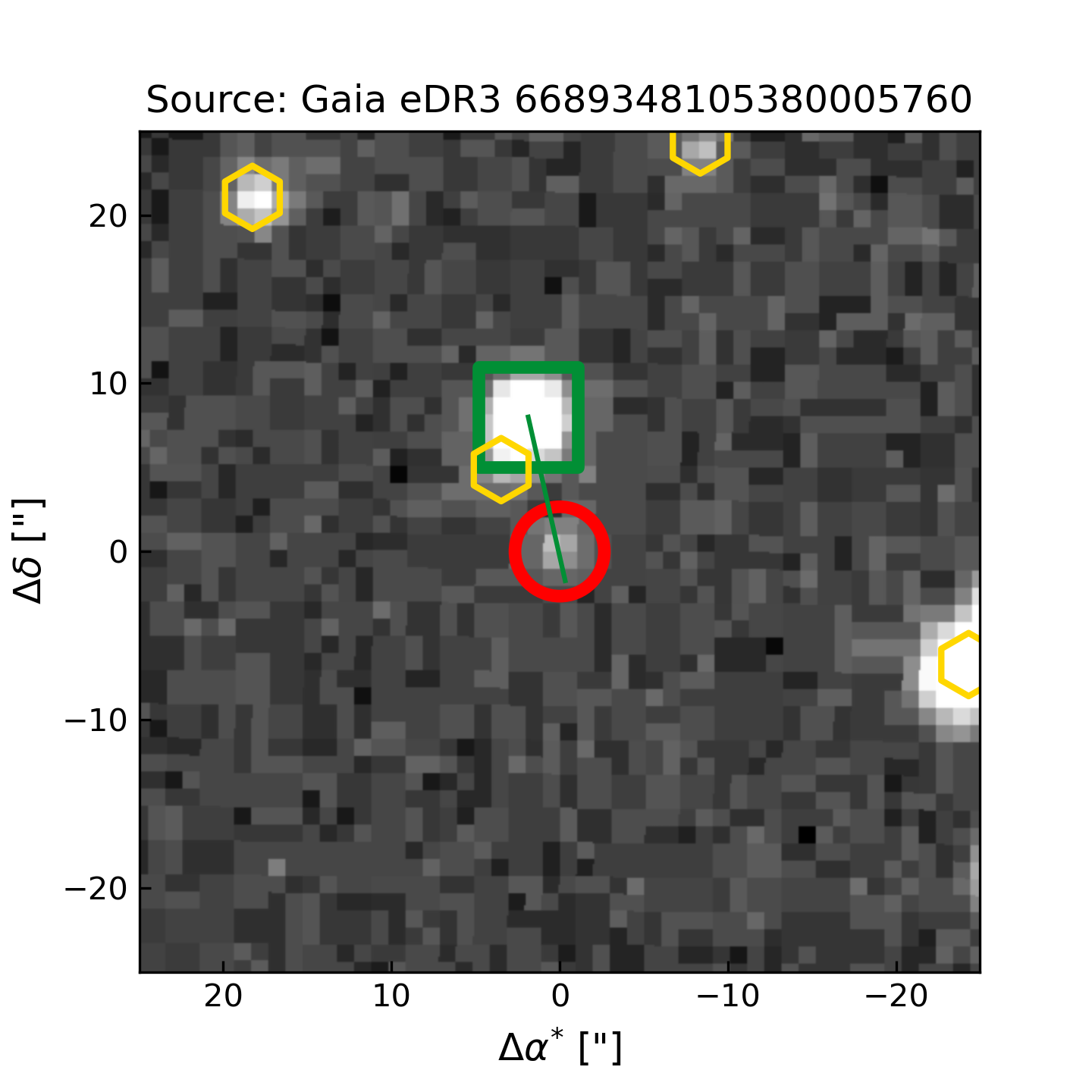}{0.24\textwidth}{(k) DSS RED}
    \fig{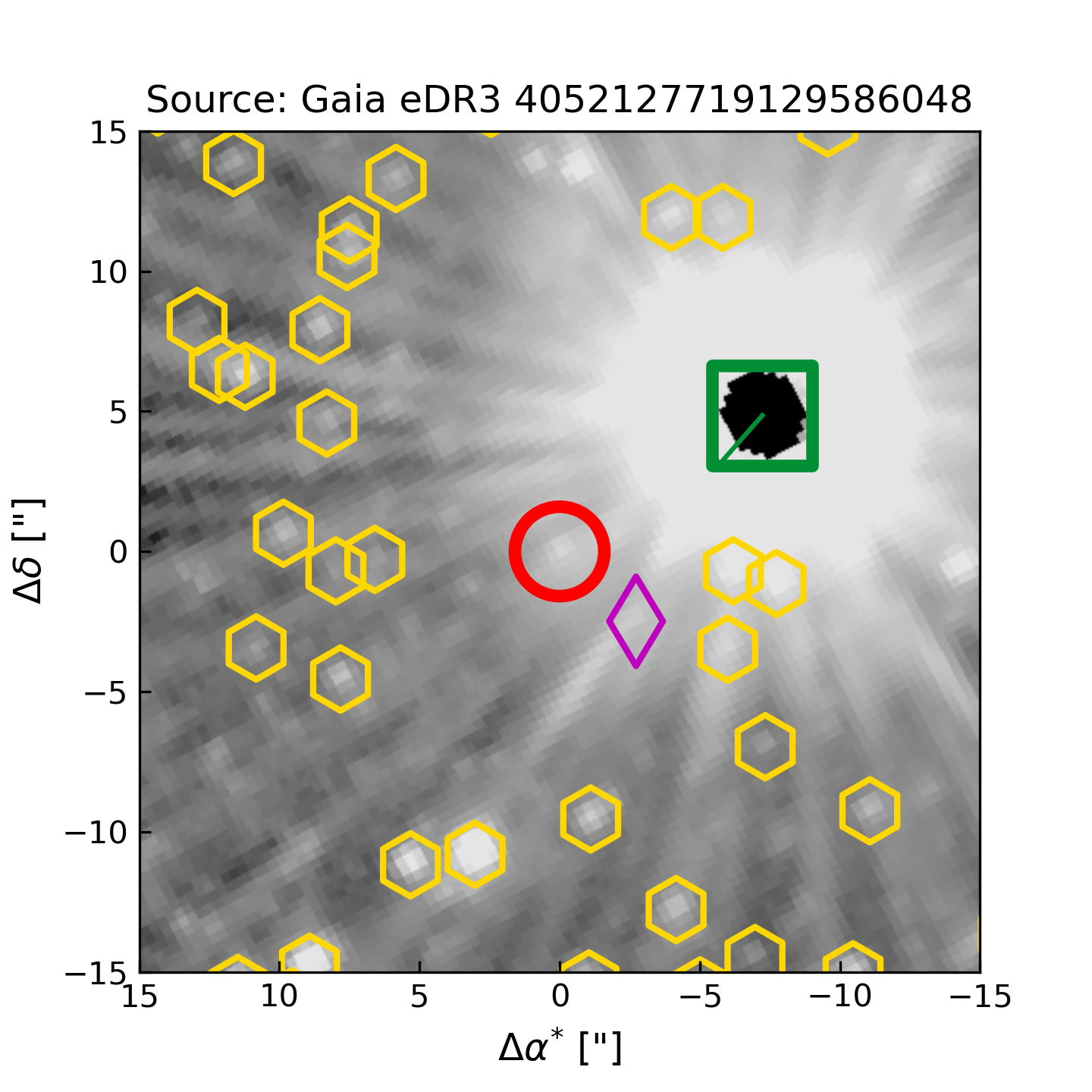}{0.24\textwidth}{(l) VVV - H }
}

\caption{Cutouts from  DSS, PanSTARRS, 2MASS and VVV used in the visual inspections. The sources are indicated as red circles and the lens as green squares epoch.
All closeby \eDR sources are marked by a yellow hexagon. 
If \Gaia list a proper-motion, we show the positions at the epoch of the image (\({\sim}2012.5\) for PanStarrs, \({\sim}1990.0\) for DSS, \({\sim}2015.75\) for VVV , and \({\sim}1998.5\) for 2MASS).
The first image (a) shows a ”candidate” which was identified as binary star using the displacement between \DRtwo (black cross) and eDR3 (yellow cross).
The Gaia DR2 positions are shown as a black cross. 
We plotted the ”BGS” also as square, and use the displacement to determine the position at the image epoch \({\sim}2013\). 
The other two images (e \& i) in the first column are labelled as bad images since the source should be visible in the PanSTARRS or DSS images. 
The second column contains images where no we were not able to confirm or rule out the existence of the background source. 
These are either highly blended by or within the saturated core of the lens stars (b \& f), to faint to be seen in DSS image (j).
Further, in some cases, the field is highly crowded to identify the source. 
For all three cases, we removed the event from our results. 
In the third column, we show examples where the source could be identified. Those are contained in our result. 
Finally, in the fourth column, we show a comparison between DSS (d), PanSTARRS(h) and VVV (l) for the same microlensing event. 
The BGS is visible in the VVV’s H band images but blended by the HPMS in the DSS and PanSTARRS images.
The Gaia sources do perfectly align with sources or features seen in the VVV image even though they are in the saturated areas of the PanStarrs and DSS image. 
There is a second candidate marked by the magenta diamond, which is aligned with a spike of the lens and hence was removed.}
\label{FIG:visiual_inspections}
\end{figure*}

These criteria can only be used when a proper motion is given in \eDR. However, a visual inspection (originally required due to reasons mentioned in subsection~\ref{SubSubsection:PSI}) has shown that most bright background sources without a five-parameter solution are within a small angular separation to an even brighter star, which results in a limited performance of \Gaia. An example is shown in panel (a) of  Fig.~\ref{FIG:visiual_inspections}.
For many of these supposed BGS, we observed that the difference between the \DRtwo and \eDR positions is similar to the difference between the positions of the HPMS, and that they are therefore co-moving stars.  
To distinguish between co-moving stars and true candidates, we used this positional offset to determine an approximate proper motion. 

Using the DR2 neighbourhood catalogue \cite{refId4}, we searched for matched sources. For \({\sim} 185,000\) of our BGS, we found at least one entry in this catalogue. To ensure a correct match we applied the following steps. Some sources have multiple entries in this DR2-neighbourhood catalogue. For those, we only considered the eDR3-DR2 pair with the smallest angular distance. We then selected only matches where the angular distance (\(\Delta\phi\)) is \(\Delta\phi < 400\mas\) 
and the magnitude difference is 
	\(\lvert\Delta G\rvert <  0.3\MAG \cdot (\Delta\phi/1\mas)^{0.2}\) 
	(see Fig.~\ref{Fig:DR2Match}).
We also excluded 1061 sources where we found a \DRtwo source within an angular distance less than \(400 \mas\), but with a magnitude difference larger than the above limit. Typically, these are brighter in \DRtwo. 
We found that for several candidates a star can be identified at the listed position, but much fainter than the listed $G$ magnitude in \DRtwo. Hence the reason for the disagreement might be an incorrect $G$ magnitude in \DRtwo.

\begin{figure}
\center
\includegraphics[width=0.95\linewidth]{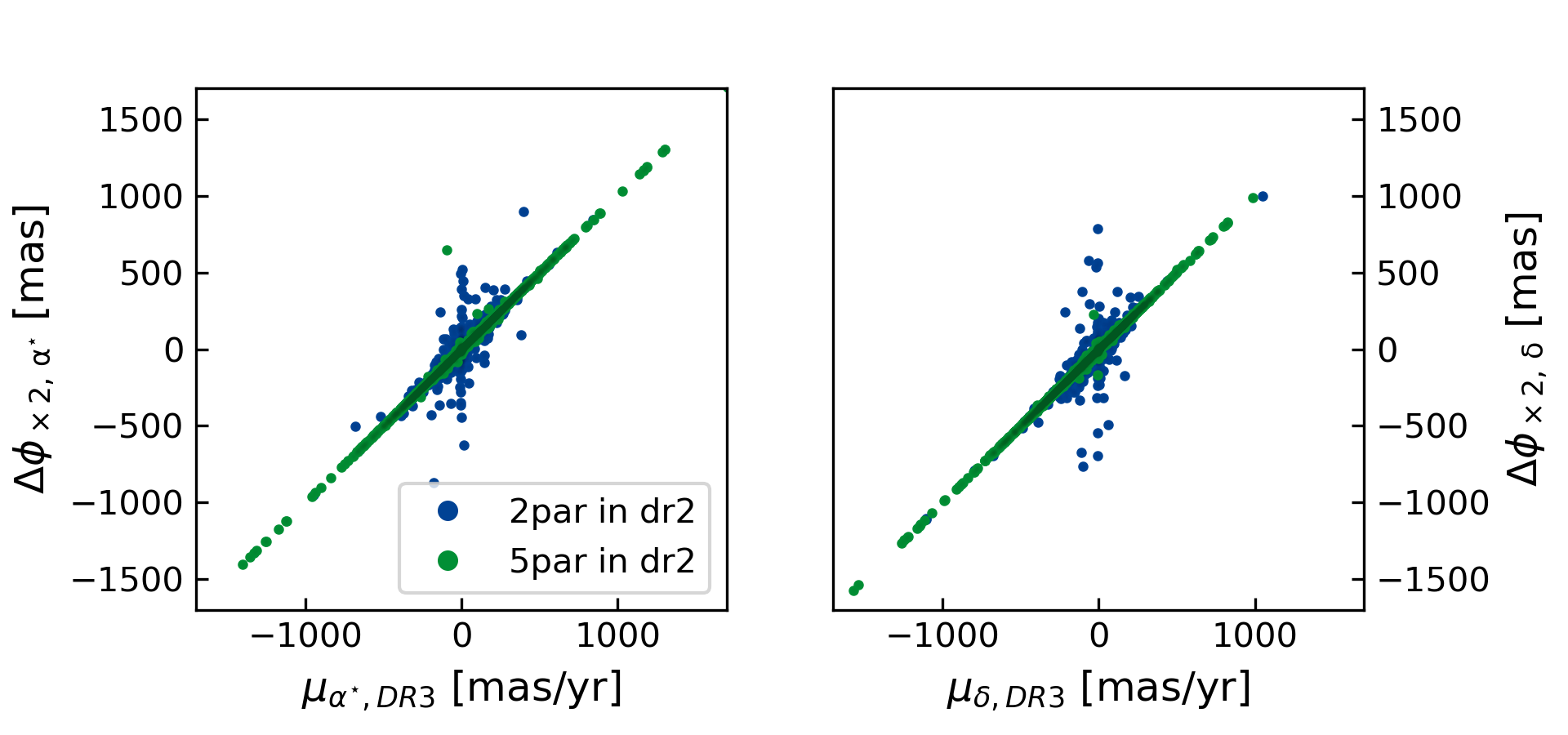}
\caption{Approximate proper motion from the position difference \DRtwo minus eDR3 versus the actual proper motion given in \eDR.  For most BGS these two are in a good agreement, especially if a 5-parameter solution exists in \DRtwo (green dots). 
The few dozen candidates with large discrepancy are most likely mismatches.
}
\label{Fig:DR2pm} 
\end{figure}

We calculated an estimate of the proper motion from the positional displacement between \eDR and \DRtwo, that is 
\begin{equation}
	 \Sou \boldsymbol{\Delta \phi}_{\times2} = 
	\left(\begin{pmatrix}  \Sou\alpha_{\mathrm{DR3}} \\  \Sou\delta_{\mathrm{DR3}} \end{pmatrix} 
	- \begin{pmatrix}  \Sou\alpha_{\mathrm{DR2}} \\  \Sou\delta_{\mathrm{DR2}} \end{pmatrix}\right) 
	 \cdot \begin{pmatrix} \cos{ \Sou\delta_{\mathrm{DR3}}} \\ 1 \end{pmatrix} /\Delta t
\end{equation}
where $\Delta t$ is the difference between the catalogue epochs of \eDR and \DRtwo (0.5~years). 
We also computed a rough estimate for the parallax as follows:
\begin{equation}
 \Sou\varpi_{\mathrm{approx}} =4.74\frac{\mathrm{km/s}}{\mpy}\cdot \frac{\lvert \Sou\Delta\phi_{\times2}\rvert/1\mathrm{yr} }{ v_{\mathrm{tan}}}\mas, 
\end{equation}
where we use a typical tangential velocity of \(v_{\mathrm{tan}} = 75 km/s\).

If the proper motion of the BGS is not given in \eDR, we used \(\Sou\boldsymbol{\Delta\phi_{\times2}}\) and  \(\Sou\varpi_{\mathrm{approx}}\)  as an estimate for the proper motion and parallax of the BGS, with an assumed standard error of  \( \Sou\sigma_{\Delta \phi_{\times2}\mathrm{,\,ra/dec}} = 5\mas\) and \( \Sou\sigma_{\varpi_{\mathrm{approx}}} = 3\mas\), respectively. These standard errors roughly reflect the half width of the distributions of all BGS with a five-parameter solution, for the corresponding parameter.
Finally we applied the same criteria as those we had for BGS with five-parameter solutions (i.e Equations (\ref{Eq:simlimit1}) -  (\ref{Eq:simlimit2})).

For candidates where neither \eDR lists a proper motion nor a partner could be found in \DRtwo we set the parallax and proper motion to zero. 
These cases were also visually inspected. 
We excluded 52  out of the 279 inspected candidates since we could not identify the BGS in any Pan-STARRS, DSS2\footnote{Digitised Sky Survey, \url{http://archive.eso.org/dss/dss/}},  2MASS \citep{2006AJ....131.1163S}\footnote{Two Micron All Sky Survey , \\ \url{https://irsa.ipac.caltech.edu/Missions/2mass.html}}  or VVV Image. 
We note that the large majority of candidates were only excluded because the source is blended in all of the considered surveys. We draw this conclusion from the fact that two thirds are located at a declination less than \(-30^{\circ}\) and outside of the Pan-STARRS footprint, while only two fifths of all inspected candidates are below \(-30^{\circ}\), and most of the inspected candidates are too faint to be identified in a DSS2 or 2MASS image (compare the images in the right column in Fig.~\ref{FIG:visiual_inspections}). Further, we note that we only inspected candidates which passed all criteria and would lead to an measurable effect.

\subsubsection[Selection on the PSI value]{Selection on the \(\Psi\) value}
\label{SubSubsection:PSI}

\begin{figure}
\center
\includegraphics[width=0.7\linewidth]{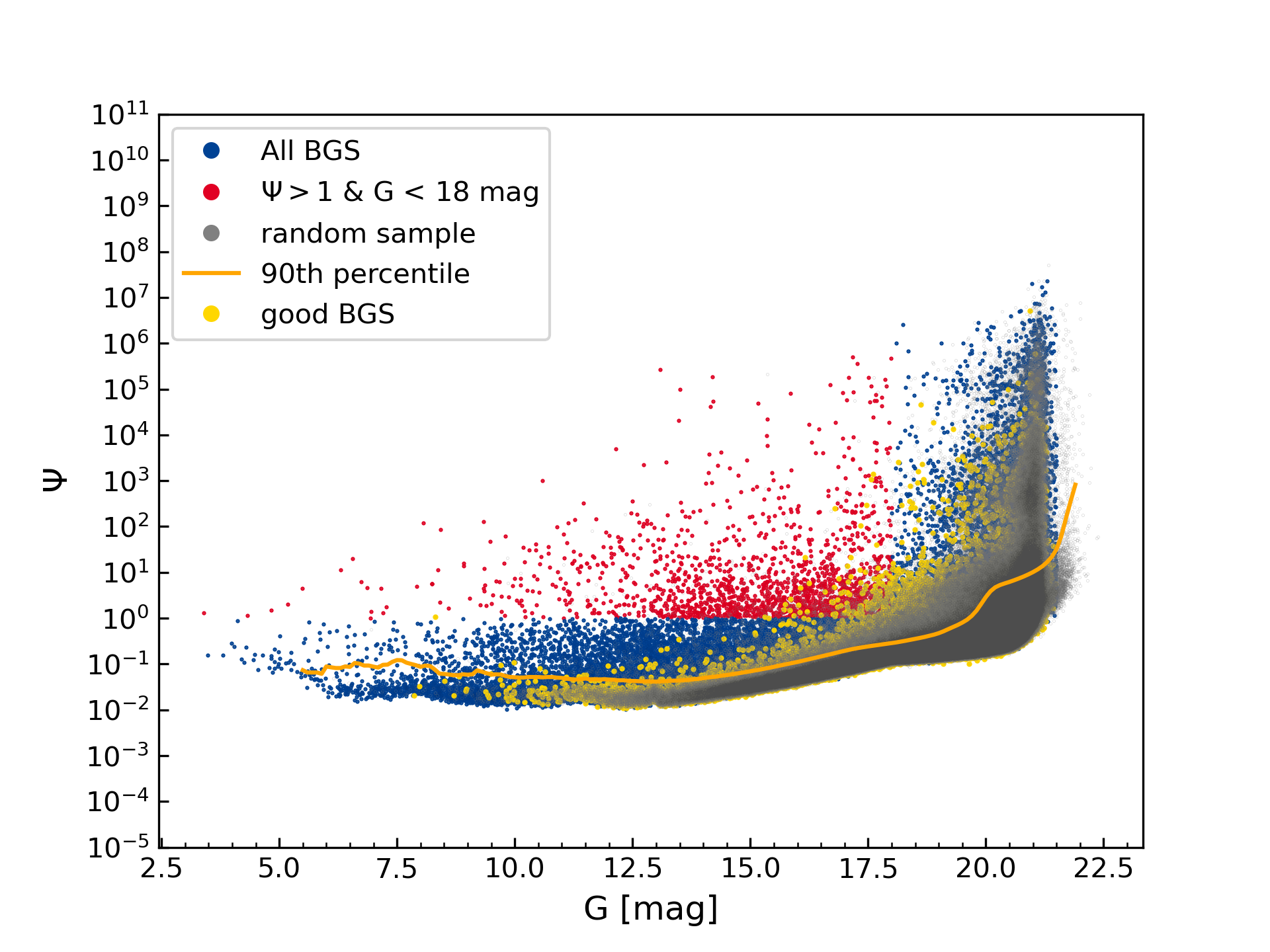}
\includegraphics[width=0.7\linewidth]{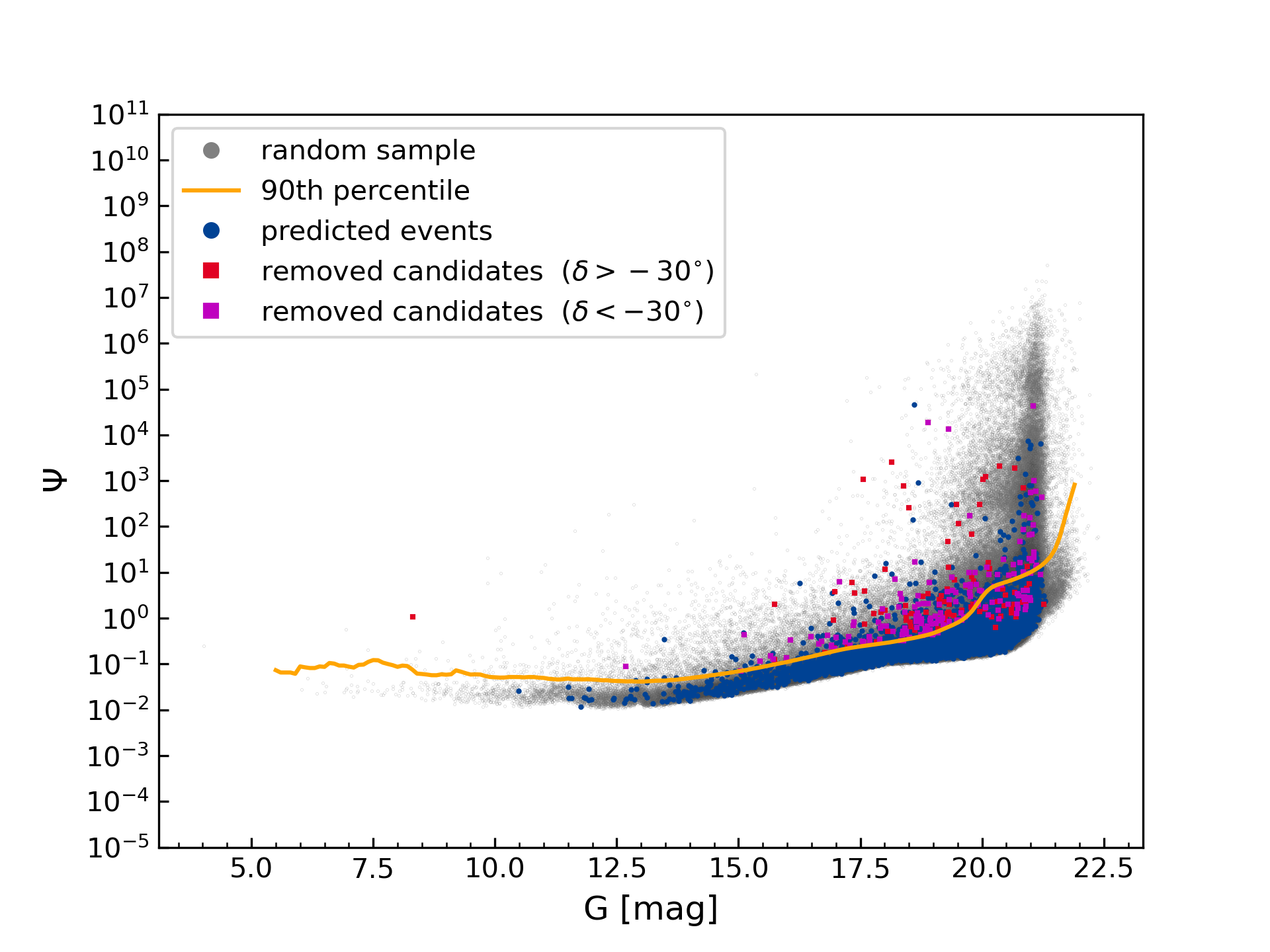}
\caption{\(\Psi\) vs. $G$ magnitude. Top: Showing the distribution for all potential BGS. BGS which would not pass the criterion defined by \cite{McGill2020} are shown in red. BGS which had passed all of our selection criteria are plotted in yellow. Only a few dozen of those  would not have passed the criterion of \cite{McGill2020}.
A sample of randomly selected stars is shown in grey and \(90\%\) of those are below the orange line. We note that the our sample has slightly larger \(\Psi\)~values compared to the random sample.  
Bottom: In blue the BGS of all predicted events are shown.
The ones excluded after the visual inspection are plotted as red (\(\delta > -30^{\circ}\)) and magenta (\(\delta < -30^{\circ}\)) squares.
The excluded candidates below the orange line were inspected since \eDR does not provide proper motions and they were not found in DR2.  
The grey dots indicate the random sample as for the top panel, with the 90th percentile shown as orange line.
}
\label{Fig:PSI} 
\end{figure}

	For sources brighter than \(G = 18\MAG\), \cite{McGill2020} proposed to use
\begin{equation}
\Psi = \frac{\sigma\_5d\_max}{1.2\mas\cdot\gamma(G)} < 1 
\end{equation}
as an additional criterion, to exclude spurious data, where  \(\sigma\_5d\_max\) is the square root of the largest singular value of the \(5\times 5\) covariance matrix of the astrometric parameters, and
\begin{equation}
\gamma = \max\left[1, 10^{0.2(G-18)}\right]
\end{equation}	
The distribution of \(\Psi\)-values as function of $G$ magnitude is shown in Fig.~\ref{Fig:PSI}.
In our raw sample (i.e.~if none of our exclusion criteria is applied), 2071 candidates would not fulfil this criterion. Out of those, 1545  violate either our criterion on the \(\Sou GoF/\sqrt{\Sou N}\) (1450)  or show a bad match between DR2 and DR3 (1468). 
The comparison with  \DRtwo has shown that a large fraction of those are most likely physical binary stars, which leads to the selection described above. Further 98 candidate pairs are excluded due to the displacement between \eDR and DR2

 Instead of a strict criterion, we visually inspected the Pan-STARRS, 2MASS and DSS2 images for 562 of the candidates where the background stars show a \(\Psi\) larger than the 90th percentile of the stars in a random sample, with similar magnitudes (shown as orange line in the top panel of Fig.~\ref{Fig:PSI}).  This leads to the exclusion of 210 candidates. Example of the visual inspection are shown in Fig.~\ref{FIG:visiual_inspections}.
 The bottom panel of Fig.~\ref{Fig:PSI} shows the \(\Psi\)-value for our predicted events in blue, the excluded candidates are plotted as red or magenta squares. 
We note that for most of the events a clear falsification was not possible due to the same reasons as mentioned in Section~\ref{SubSubsection:DR2pm}.

\subsection{Approximate Masses and Einstein Radii}
\label{subSection:Mass}
\begin{figure}
\center
\includegraphics[width=0.95\linewidth]{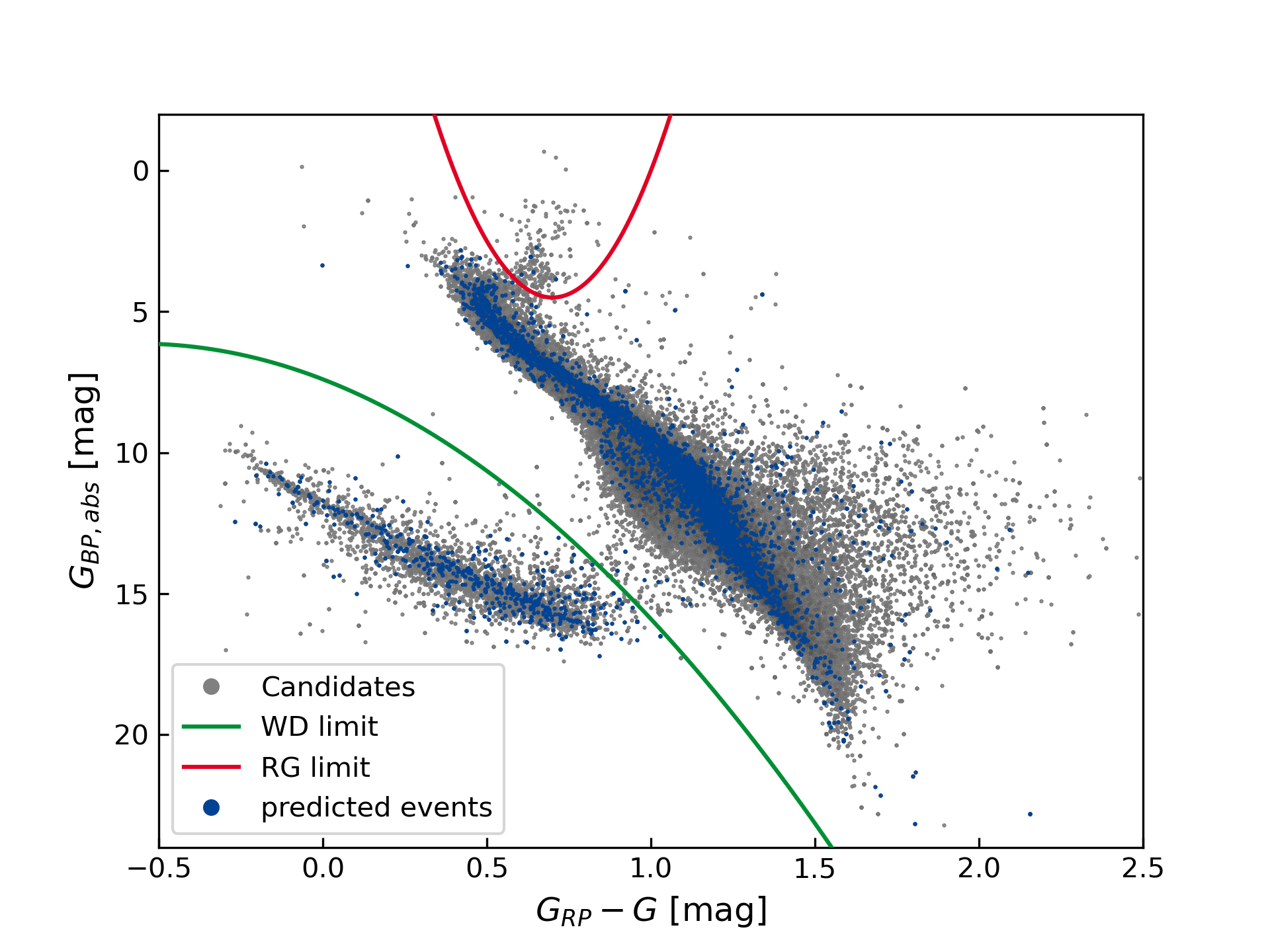}
\caption{Colour-magnitude diagram of all potential lenses (grey) with full \eDR photometry \(G\), \(G_{RP}\), and \(G_{Bp}\). The blue dots indicate the lenses of the predicted events. All stars above the green line are considered as red giants and all sources below the red line as white dwarfs.}
\label{Fig:CMD} 
\end{figure}

For each of the potential astrometric-microlensing events, we determined the Einstein radii based on an assumed lens mass.  We used this to determine an expected deflection and magnification of each event. We distinguished between main-sequence stars (MSs), white dwarfs (WDs) and red giants (RGs), using the following limits (see Fig.~\ref{Fig:CMD}):
 
\begin{align} 
WD:\,\,\, & G_{BP,abs} > 4.0 \cdot (G - G_{RP} )^{2} + 4.5 \cdot (G - G_{RP}) + 7.4 \\ 
RG:\,\,\, & G_{BP,abs} < -50.0 \cdot (G - G_{RP} )^{2} + 70.0  \cdot (G - G_{RP} ) - 20.0
\end{align} 
where  \(G_{BP,abs}\), is  the absolute \(G_{BP}\)  magnitude determined using the distance modulus. These limits were slightly adjusted compared to \Kt.
HPMS without a \(G_{BP}\) and \(G_{RP}\) magnitude given in \eDR are assumed to be main-sequence stars. 

For WDs and RGs we assumed typical masses of
\(
M_{\rm WD} = (0.65 \pm 0.15) M_{\odot}\)
and \(M_{\rm RG} = (1.0 \pm 0.5) M_{\odot}\),
respectively, and for MSs  we determined the mass based on the absolute \(G\) magnitude (\(G_{abs}\)), using the following equation based on the mass-luminosity function as determined in \Kt (Section 3.4): \\
for \(G_{abs} < 8.85:\)
\begin{equation} 
\qquad\ln\left(\frac{M}{M_{\odot}}\right)   = 0.00786\,G_{abs}^{2} -0.290\,G_{abs} + 1.18
\end{equation}
for \(8.85 <  G_{abs} < 15: \)
\begin{equation} 
\qquad \ln\left(\frac{M}{M_{\odot}}\right) =         -0.301\,G_{abs}  + 1.89, 
\end{equation}
For MSs we assumed an uncertainty of \(10\%\).
Stars with \(G_{abs} > 15\) lie in the Brown Dwarf (BD) regime, where the mass-luminosity function is not valid.
For these, we assumed a mass of
\begin{equation} 
M_{BD} = (0.07 \pm 0.03)\Msun 
\end{equation}

We note that this is only a rough estimate. 
For example it treats sub dwarfs like MS Stars. 
However, measuring the mass is aim of the observations of the astrometric microlensing event.

Using Eq.~(\ref{EQ:ThetaE}) we then determined the Einstein radii, using the masses determined above and the \eDR{} parallaxes of lens and source. For the source parallaxes, we also used the prior derived from the DR2-eDR3 displacement, if only a two-parameter solution is provided by \eDR{}.

\subsection{Forecasting the motion}
\label{subSection:FindClosest}
For about \(136\,000\) candidates, we searched for the closest approach between source and lens. Using elementary geometry, we first determined an approximate minimum angular separation \((d_{\rm CA,\, approx})\) and approximate epoch of the closest approach (\(T_{\rm CA,\,approx}\)):
\begin{equation}
d_{\rm CA,\,approx} =  \left\lvert\frac{\Delta\alpha^{\ast} \cdot \Delta\mu_{\delta} - \Delta\delta \cdot  \Delta\mu_{\alpha^{\ast}}}{\sqrt{\Delta\mu_{\alpha^{\ast}}^{2}+ \Delta\mu_{\delta}^{2}}}\right\rvert
\end{equation}
\begin{equation}
T_{\rm CA,\,approx} =  -\frac{\Delta\alpha^{\ast} \cdot \Delta\mu_{\alpha^{\ast}} + \Delta\delta \cdot  \Delta\mu_{\delta}}{\Delta\mu_{\alpha^{\ast}}^{2}+ \Delta\mu_{\delta}^{2}} + 2016.0
\end{equation}
where \(\Delta\alpha^{\ast}\), \(\Delta\delta\), \(\Delta\mu_{\alpha^{\ast}}\), \(\Delta\mu_{\delta}\) are the differences in positions and proper motions between lens and source in right ascension and declination. Note that this approximation neglects the periodic motion caused by the parallax difference. 

For the next step, we only considered the \({\sim}16,000\) candidates where the predicted shift of the major image is larger than \(0.03 \mas\) or the approximate minimum angular separation is less than two times the parallax difference. 
For those we searched for the exact closest approach. We determined the position of lens and source in equatorial Cartesian coordinates as a function of time (\(t\)) after the \Gaia reference epoch \(t_{0} = J2016.0\),
\begin{equation}
\boldsymbol{x}(t) = \begin{pmatrix} 
\cos(\delta+\mu_{\delta}\,t)\cdot \cos(\alpha+\mu_{\alpha}\,t)\\
\cos(\delta+\mu_{\delta}\,t)\cdot \sin(\alpha+\mu_{\alpha}\,t)\\
\sin(\delta+\mu_{\delta} t)
\end{pmatrix}	\frac{1000\mas\cdot pc}{\varpi} + \boldsymbol{E}(t).
\end{equation}
where \( \boldsymbol{E}(t)\) is the geocentric location of the Sun at \(t_{0} + t\), in equatorial Cartesian coordinates. We note that \(\mu_{\alpha} = \mu_{\alpha^{\ast}}/\cos(\delta)\). This is retrieved from the astropy packages \citep{astropy}. 
Further, we determined the angular separation between lens and source via:  
\begin{equation}
d(t) = 2\cdot\arcsin\left(\left|\frac{\boldsymbol{x}(t)}{\lvert \boldsymbol{x}(t) \rvert} -\frac{\Sou \boldsymbol{x}(t)}{\lvert \Sou \boldsymbol{x}(t)\rvert}\right| \cdot 0.5\right).
\end{equation}
We then evaluated this function on a one-week grid within two years around the previously determined approximate epoch. By comparing consecutive grid points, we detected local minima and then applied a nested-intervals algorithm to determine the precise epoch and minimum distance for each local minimum.
As edges (\(t_{0}\), \(t_{3}\)) of the starting interval we use the two adjacent neighbours of the previous found minimum minimum, and split it according to the golden ration at  \(t_{1}=t_{0}+38.2\%\cdot(t_{3}-t_{0})\), \(t_{2}=61.8\%\cdot(t_{3}-t_{0})\) of the interval length. We then shrink the interval such that in each step the epoch for the minimum of \(d(t_{1}), d(t_{2})\) stays in the center of our interval (the new intervals is \(t_{0}-t_{2}\) or \(t_{1}-t_{3}\), respectively), until the width is less than \(3\cdot10^{-10} yr\approx{10\mathrm{ms}}\), i.e much smaller than the expected error.
Finally, by comparing all detected local minima, we selected the global minimum.

\subsection{Computing the expected astrometric effects}
\label{subSection:effect}

Using equations  (\ref{EQ:shiftp}) -  (\ref{EQ:shiftlum}) and (\ref{EQ:mag}), we computed the expected astrometric shifts for the major image (\(\delta\theta_{+}\)), for the centre of light (\(\delta\theta\)), and for the centre of light including luminous-lens effects (\(\delta\theta_{\mathrm{lum}}\)), as well as the expected photometric magnification. We determined the uncertainties of these predictions using a Monte Carlo approach, where we picked 5000 realisation. The results corresponds to the median and the upper and lower errors to the 
\(15.87th-50th\) and \((84.13th-50th\)  percentile, respectively.  
We do not include any co variances between the different input parameters.

Finally, we only selected predicted events with an expected shift of the major image larger than \(\delta\theta_{+} = 0.1\mas\) or with an expected magnification larger than \(\Delta m = 1\,\mmag\). 

Since several high-precision astrometric instruments like \Gaia or JWST are or will be located at the Sun-Earth Lagrange point L2, we repeated our calculations with a 
\(1\%\) larger parallax. These results are labelled in the online table with ``L2\_'' as a prefix. However, the epochs and expected separations differ only very little.

\section{Results: Predicted astrometric-microlensing events}
\label{Section:Results}
\begin{figure*}
\center
\includegraphics[width=0.95\linewidth]{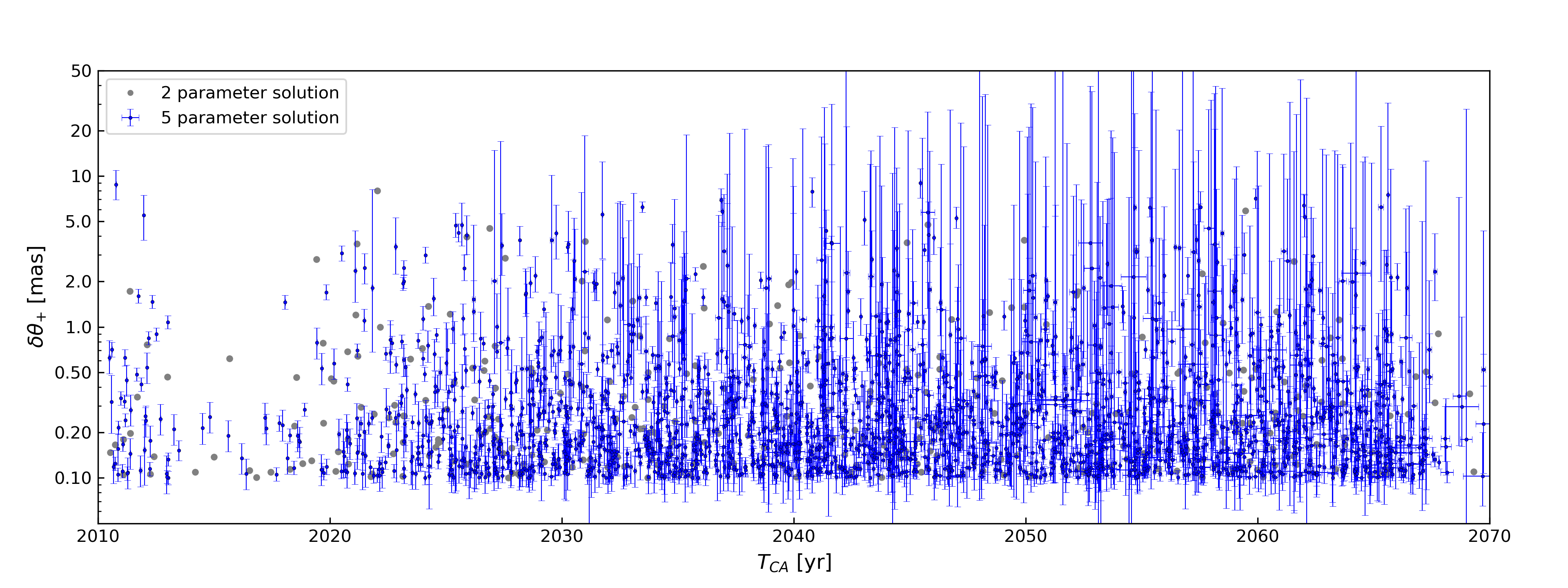}
\includegraphics[width=0.95\linewidth]{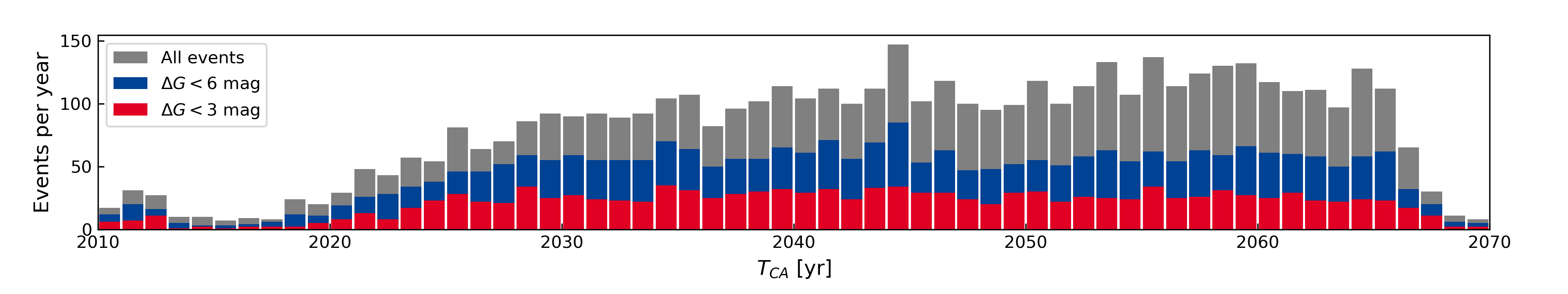}
\caption{Top: Expected maximum shifts for all astrometric-microlensing events with a magnitude difference \(\Delta G <6 \MAG\). 
The grey dots indicate the events where the proper motion and parallax of the source is unknown. The blue dots show the events with a five-parameter solution for the background source, as well as the determined standard errors. On the x-axis the epoch of closest approach $T_{\rm CA}$ is displayed. The apparent paucity of events during the \Gaia mission time is due to the angular resolution limit of \eDR.
Bottom: Number of predicted events per year. The red bars shows the events with \(\Delta G < 3\MAG\),  the red + blue bars show the events with \(\Delta G <6\MAG\), and the red + blue + grey shwos all predicted events. }

\label{Fig:Result} 
\end{figure*}

We report the prediction of 4842 astrometric-microlensing events based on \eDR, caused by 3791  distinct High-Proper-Motion Stars. 
The relevant date on the individual events are listed in the online table. This can be accessed through the GAVO Data Center\footnote{German Astrophysical Virtual Observatory,\\\urlresult}, and is
also available through the TAP service\footnote{\urltab}
or through the Virtual Observatory (look for ``Astrometric-Microlensing Events Predicted from \Gaia eDR3'')
A description of all the coloumns can be find in Table \ref{tab:column_description} + \ref{tab:column_description_continue} 

The table contains updated parameters for 3084 events already predicted from \DRtwo (see subsection \ref{subsection:OldResults}). For those we provide updated and more precise information on the separation and epoch of the closest approach. The 1758 newly predicted events are mainly due to the reduced lower limit for the proper motion of the HPMS (\(\mu_{tot}>100\mpy\) instead of \(\mu_{tot}>150\mpy\) in \Kt;
see subsection \ref{subsection:NewResults}).

A total of 869  events predicted in \Kt could not be re-discovered with \eDR, mainly due to the  stricter quality requirements.  
For about 200 of those events, \eDR led to strong evidence that lens and source are co-moving stars, based either on the proper motion given in \eDR or on the positional displacement between \eDR and \DRtwo.
 
\subsection{Updated results from \Gaia DR2}
\label{subsection:OldResults}
In \cite{2018A&A...615L..11K} and \Kt we predicted 3908 astrometric-microlensing events, and \cite{2018AcA....68..183B} and \cite{2018A&A...618A..44B} predicted 79  and 2509 further events, respectively.
These partly overlap, so that a total of 5766  events had been predicted from   \DRtwo. 
In this section we investigate which of the events are found within our new results, and why certain events do not show up within our new sample.
In  the rest of this section the  numbers are given for the sample of the 3908 events which we predicted with \DRtwo, followed in parentheses by the  number for the full sample of 5766 \DRtwo predictions .

\subsubsection{Match between {\it Gaia} DR2 and eDR3}
In order to match the source IDs between \eDR and \DRtwo, we again used the DR2 neighbourhood catalogue provided with \eDR. If in \eDR multiple sources are provided for a single \DRtwo source, we only considered those with the smaller angular separation as a genuine match. Source pairs with angular separations below \(450\mas\) and with magnitude differences below \(0.5\MAG\) are labelled  as good matches, the others as bad matches. For each of the HPMS we were able to detect a counterpart in \eDR. Among those, 45 (45)  were labelled as bad matches due to large angular separations and 3 (3) due to a missing $G$ magnitude in \eDR. 
For 5 (6)  of the  BGS, the DR2 neighbourhood catalogue does not list any entry. 
For 3 out of the 6, the BGS is clearly visible in the Pan-STARRS images, and for one of them a slight distortion of the PSF can be observed in VVV. For the other 2, the images were strongly blended  by the HPMS. 
Additionally, 114 (140)  BGS show a bad match between \DRtwo and \eDR (\(d> 450mas: 26\,(39) \), \(\Delta G > 0.5mag: 90\,(103)  \); \(G_\text{DR3} = None: 18\,(31)   \)).
In total, for 166 (193)  events, it was not possible to find a good match for the BGS and HPMS.  
Further, for 21 (33) of those, the BGS and HPMS are matched to the same \eDR source. 
This is also true for 2 events where both matches seem to be good. 
Additionally, two BGS are matched to the same \eDR source, but one of those matches is classified to be bad.
Finally, the lenses of 26 (860) and 81 (82)  events, are not classified as HPMS in \eDR, due to their proper motion being less than \(100\mpy\) or because no proper motion is given, respectively,
and for 21 (806) events the background source is outside of our search rectangle. 
The high numbers for  the full sample are due to the extended search by \cite{2018AcA....68..183B} for events until the year 2100, and no cuts on the proper motion. 
In conclusion, within the search windows around individual HPMS we were able to find 3755 (3990)  events among our original 3908 (5766)  candidates before the quality cuts.

\subsubsection{Excluded events}

This subsection gives the statistics of the various reasons for excluding some of the predicted events.  We note that an event could be excluded due to multiple reasons.

	Out of the 3755 (3990)  events described in the previous subsection, 402 (404) events were excluded because the HPMS did not pass all quality criteria. This was mainly caused due to a \({\rm \texttt{ruwe}} > 2\) (396 (398)  events). For 7 (7)  events the HPMS did not pass the \( G_{flux}/\sigma_{G_{flux}}\cdot n_{obs}^{1.5}> 3\cdot 10^{5}\) criterion. 
We note that  \cite{2021arXiv210111641R}  determined a fidelity value less than 0.5  only for two of the excluded events. Except for one which has a value of 0.87, they have a value larger than 0.99 (the fidelity value lies between \(1.0={\rm Good}\) and \(0.0={\rm Bad}\)). 
Hence the \({\rm \texttt{ruwe}} = 2\) might be too strict a limit. 
Consequently, events which are only excluded due to a high \texttt{ruwe} might be real events.  

Based on the information of the BGS we excluded 283 (285)  events. 
These are mainly due to a high \(GoF/n > 1.24\) (197 (197)  events) which is equivalent to \({\rm \texttt{ruwe}} > 2\),
or due to a strange magnitude difference between DR2 and DR3 (207 (209)   events). 
Additionally, 18 (19)  events were excluded due to a significantly negative parallax,
and 34 (35)  events were excluded due to a standard error in the position larger than \(10\mas\). Furthermore we found 198 (200)  events where the motion of the background source in \eDR, or the positional change between \eDR and \DRtwo indicates that the supposed lens and source star are co-moving stars.
Finally, 26 (29) events were excluded after visual inspection of the BGS with an atypically high \(\Psi\) value. In addition, it is possible that some of the events found in \eDR but with an expected shift less than \(0.1\mas\) would not have passed the visual inspection. 

In the end 3039 (3264)  astrometric-microlensing events passed all of our criteria (and 3352 (3579)  when using the fidelity value instead of the \texttt{ruwe} and parallax error). 

\subsubsection{Re-detected Events}
For 150 (180) of these 3039 (3264) re-detected events, the expected effects are below our selection limit of \(\delta \theta_{+}> 0.1\mas\) or  \(\delta m> 1\mmag\). 
For 8 (20) events this is caused by a significant difference in the mass  of the lens. For the 8 events in \Kt this is due to a classification as MS instead of a WD. 
For the other 142 (150) events, the differences between the values determined from \Gaia DR2 and eDR3  are not significant (\(<3\sigma\)).

Finally, our result contains 2888 (3083) events previously predicted.
The majority of the events, 2818 (2997) have comparable angular separations. Only for 70 (86)  events we found a significant difference (\(>3\sigma\)) between \DRtwo and eDR3. These are caused by significant changes within the \Gaia Data. For the epoch of the closest approach, we also find a good agreement for most of the events. However, 181 (193) show offsets between 0.5 and 1.0 years. 
Beside the significant changes within the \Gaia Data, this is due to two local minima with roughly the same minimum separation, resulting in a switch from one to the other as global minimum. This effect is not covered in our error estimation. However, for most of those events a measurable  shift is also expected during the epoch predicted from \DRtwo.
By using \eDR the uncertainties for both epoch and position are smaller by roughly a factor of~2.

Except for 17 (78) events, the expected effects are in a good agreement, too. 
Significant differences are mainly caused by large differences in the assumed masses –– this concerns 10 (56)  events. For 7 (22) events this is caused by significant differences in the impact parameters. We do not observe a reduction of the mean errors, since these are dominated by the uncertainties of the mass.

\subsection{New events from \Gaia eDR3}
\label{subsection:NewResults}
In addition to the 2888 (3083)  previous events, we  predict 1756  new astrometric-microlensing events. This is mainly due to the reduction of the lower limit in the  HPMS proper motion. 
However, we also found 563  new events with a proper motion larger than \(150\mpy\). In most of these cases (487  events), \DRtwo did not contain the background source, and for 76 events, \eDR indicates the lens as binary star, while  \DRtwo only lists one component.

\subsection{Properties of the full sample}
Our full sample of predicted astrometric-microlensing events based on \eDR contains 4842  events caused  by 3791  distinguished HPMS. The number of events for the different criteria are summaized in Table \ref{tab:result}.
As in \Kt, these are mainly located towards dense areas in the galactic disc, or towards the Magellanic Clouds. 
Since astrometric microlensing favours close-by massive stars as lenses, roughly half (2160)  of the events have a magnitude difference larger than~6.  For 2682  events the magnitude difference is smaller than~6, and for 1246  events the lens star is at most 3~magnitudes brighter.  In the following we give the corresponding numbers for each of these two subsamples in parentheses, i.e.~the total number of events is followed by (events with \(\Delta G<6\), events with \(\Delta G<3\)).
Finally, we found 285  events where the BGS is brighter than the lens. 
We note that bright lenses tend to have either large masses or small distances, both lead to large (angular) Einstein radii. Hence, a measurable  shift is also expected at larger angular separations, where the source might  be detectable next to a bright star. 

Out of the 4842 (2682, 1246)  events,  473 (260, 111) have an expected shift larger than \(1\mas\), and for 532 (301, 135)  the expected shift is between \(0.5\mas\) and \(1\mas\). 
Especially the 260 events with \(\Delta G < 6\) and \(\delta\theta_{+}>1\mas\) are promising targets. However, also the impact parameter and observability should be considered in order to select targets for observation, which can differ between Telescopes and instruments.

In the unresolved case, i.e. when only the shift of the  centre of light can be observed, we found  393 (393, 380) events with an expected effect larger than \(\delta\theta_{\rm lum} >0.1\mas\). This can only be observed for events with low contrast between HPMS and BGS, since the blending by the HPMS strongly reduces the expected shift. All of the 50 events with an expected shift of the centre of light above \(0.5\mas\) have a magnitude difference below \(\Delta G = 3\). 
About \(40\%\) of those (147 out of 393) are caused by a WD.
We note that the contrast between HPMS and BGS, and therefore the expected shift of the centre of light depends on the flux ratio in the selected filters.

\begin{figure}
\center
\includegraphics[width=0.95\linewidth]{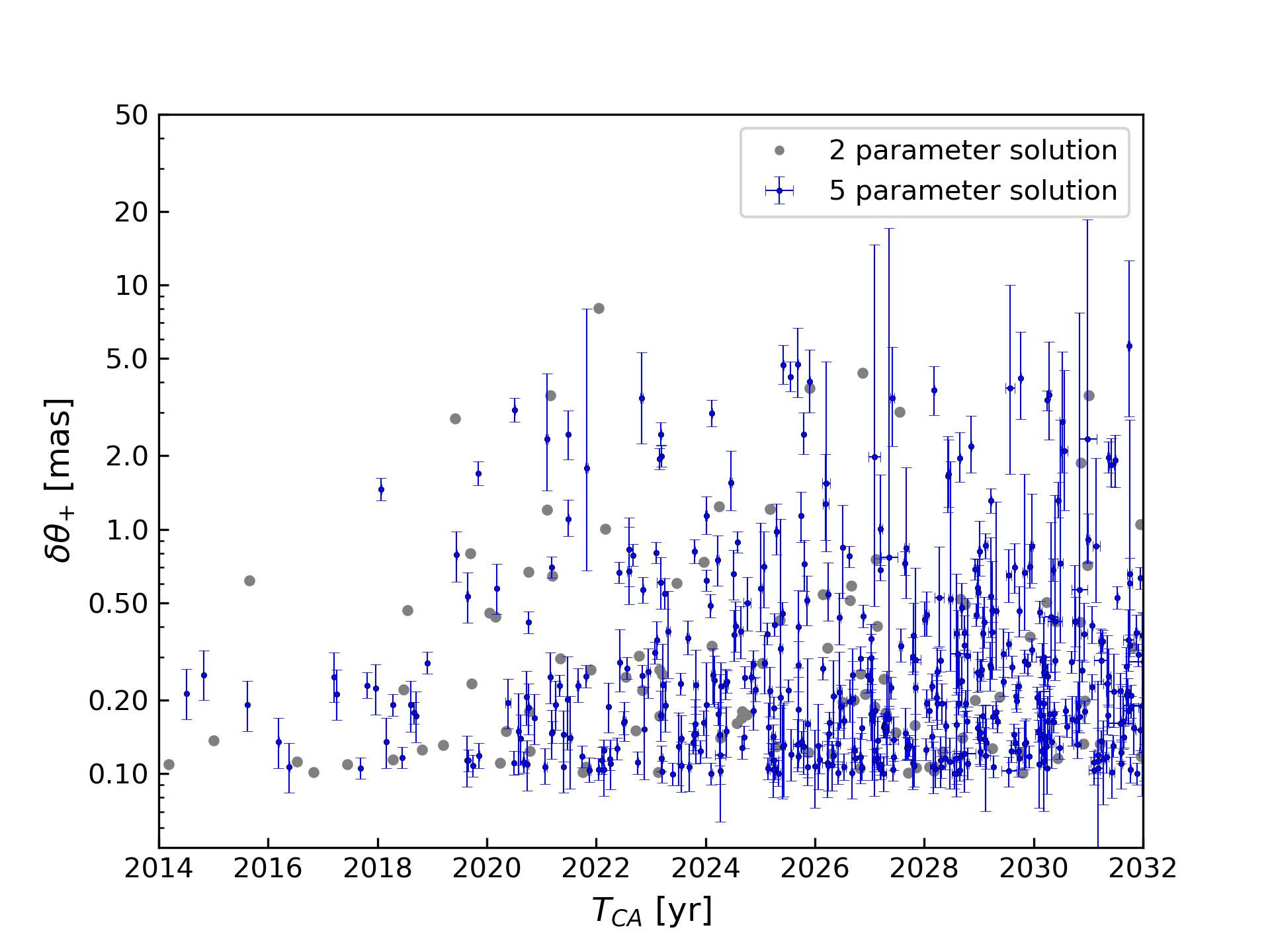}
\includegraphics[width=0.95\linewidth]{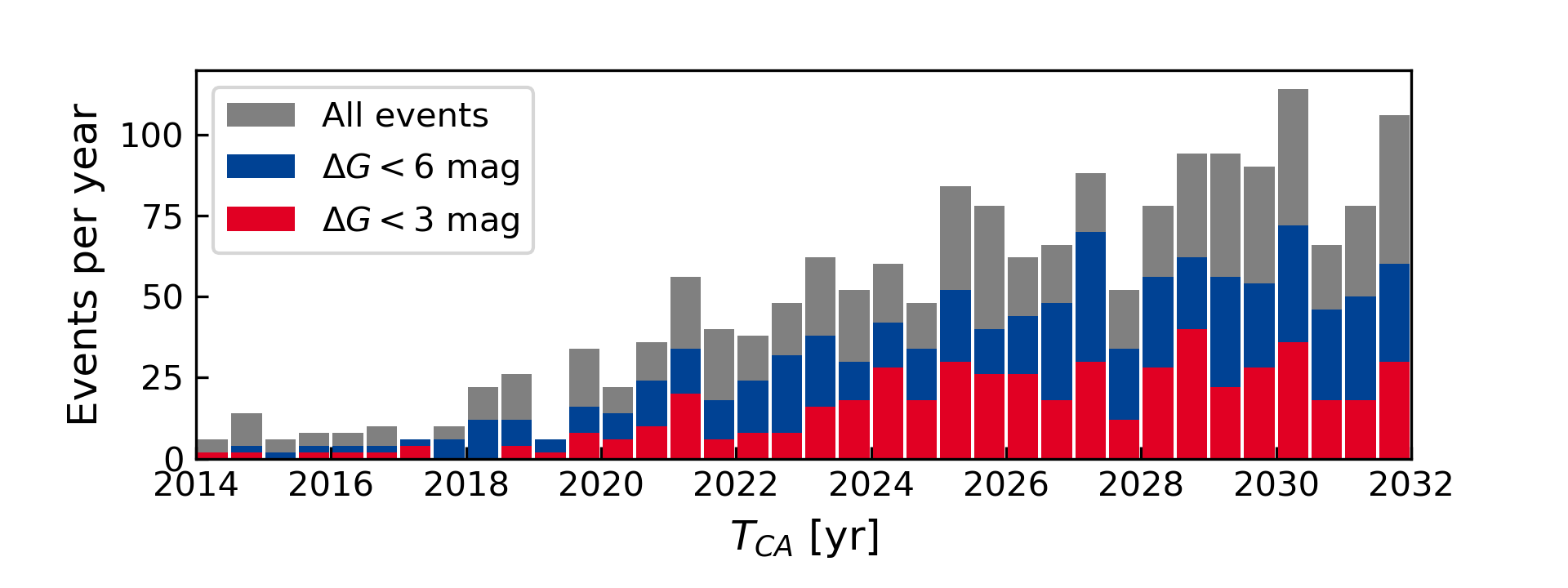}
\caption{Top: Maximum shifts for all expected astrometric-microlensing events between 2014 and 2032 and a magnitude difference \(\Delta G <6 \MAG\), with (blue) and without (grey) five-parameter solutions for the background source in \eDR. Bottom: Number of events per year (counted in half-year bins and multiplied by a factor of 2) for all events (grey + blue + red), events with a magnitude difference \(\Delta G <6\MAG\) (blue + red) and events with a magnitude difference \(\Delta G <3\MAG\) (red).}
\label{Fig:Results2030} 
\end{figure}

The expected shifts and predicted epochs of minimal separation are shown in the top panel of Fig.~\ref{Fig:Result}, and the number of events per year are shown in the bottom panel. 
The  number of events per year is 94 (52, 24). 
Until the year \({\sim} 2030\) the number of predicted events is  low. This is caused by \Gaia's limited spatial resolution and performance for close pairs. For the next decade (2021-2031) we found 685 (443, 218)  events; these and the events during the \Gaia mission are shown in Fig.~\ref{Fig:Results2030}.
The latter will become especially interesting when \Gaia shall publish its individual astrometric measurements, planned for the Data Release~4 (expected in the year 2024) and for the final data release after the extended mission (expected about 2029). 
Within the next decade the number of events per year strongly increases with time, where events caused by faster HPMS tend to be earlier.
Within the next decade we expect 76 (46, 17)  events with an expected shift larger than \(1\mas\).

\subsubsection{Time scales}
\begin{figure}
\center
\includegraphics[width=0.95\linewidth]{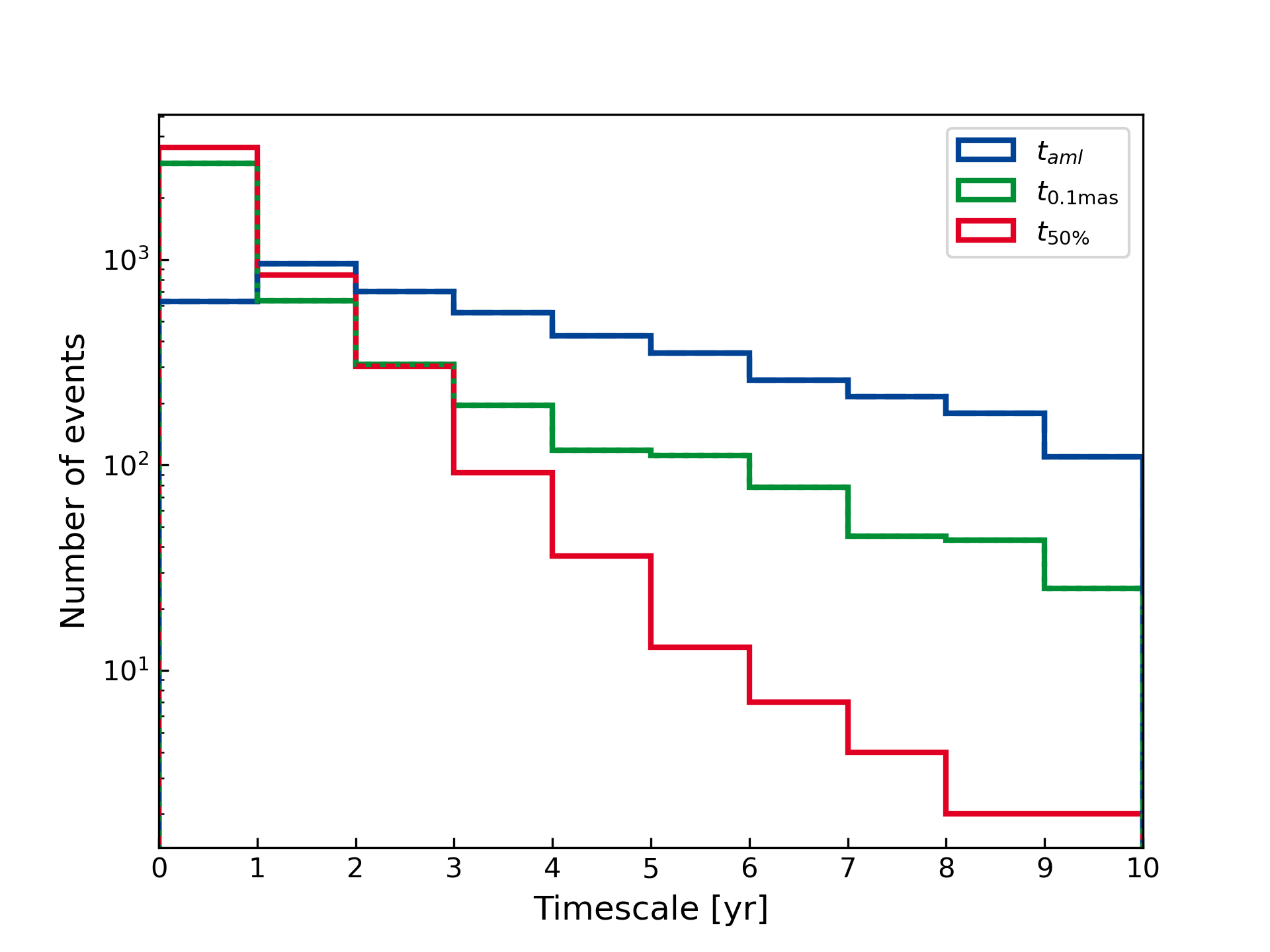}
\caption{Distribution for
\(t_{aml}\) (blue), \(t_{0.1\mas}\)(green) \(t_{50\%}\) (red). 
The trends of the distribution for \(t_{aml}\) and \(t_{0.1\mas}\) continues towards longer timescales. 
Note the logarithmic scale.}
\label{Fig:time_scales_hist} 
\end{figure}

Typically, a shift larger than \(0.1\mas\) from  the unlensed position is expected over a time range between \(1.1 \yr\) (15.87th percentile) \(7.7 \yr\) (84.13th percentile) with a median of \(3.2 \yr\). And it takes typically between \(0.037 \yr\) (\(27 \mathrm{weeks}\),  15.87th percentile) and \(3.6\yr\) (84.13th percentile) to observe a change in the positional shift by \(0.1\mas\), with a median of \(0.53 \yr\). Both distributions are shown in Figure~\ref{Fig:time_scales_hist}. For 303 events \(t_{0.1\mas}\) is between \(5 \yr\) and \(10 yr\), and for 209 events \(t_{0.1\mas}\) is longer than \(10 \yr\). The expected shifts for those events are below \(0.2\mas\) and  \(0.15\mas\), respectively.
Additionally, 39 events do not list an \(t_{0.1\mas}\), since they show large uncertainties, an the expected shift is below \(0.1\mas\) when the uncertainties are not included in the analysis.
Since all of the events are included in the \eDR, events with large \(t_{0.1\mas}\) can benefit from the existence a precise J2016.0 position. However, those are hard to observed with an accuracy of \(0.1\mas\) anyway. For 263 of the 512 "slow" events (\(t_{0.1\mas}>5\yr\)) a change in the position by \(50\% \cdot \delta\theta_{+} (\simeq 0.05-0.8\mas)\) is between \(2 \yr\) and \(5 \yr\), and for 221 shorter than \(2 \yr\).
we found a strong correlation between \(\log(t_{0.1\mas})\) and  \(\log(u^{2}/\mu_{rel})\) with a correlation coefficient of 0.98, leading to: 

\begin{equation}
    t_{0.1\mas} \simeq 0.11\yr\cdot \left(\frac{u_{min}^{2}\cdot 1\mpy}{\mu_{rel}}\right)^{1.16}
\end{equation} 

\subsubsection{Events caused by White Dwarfs}
\begin{figure}
\center
\includegraphics[width=0.95\linewidth]{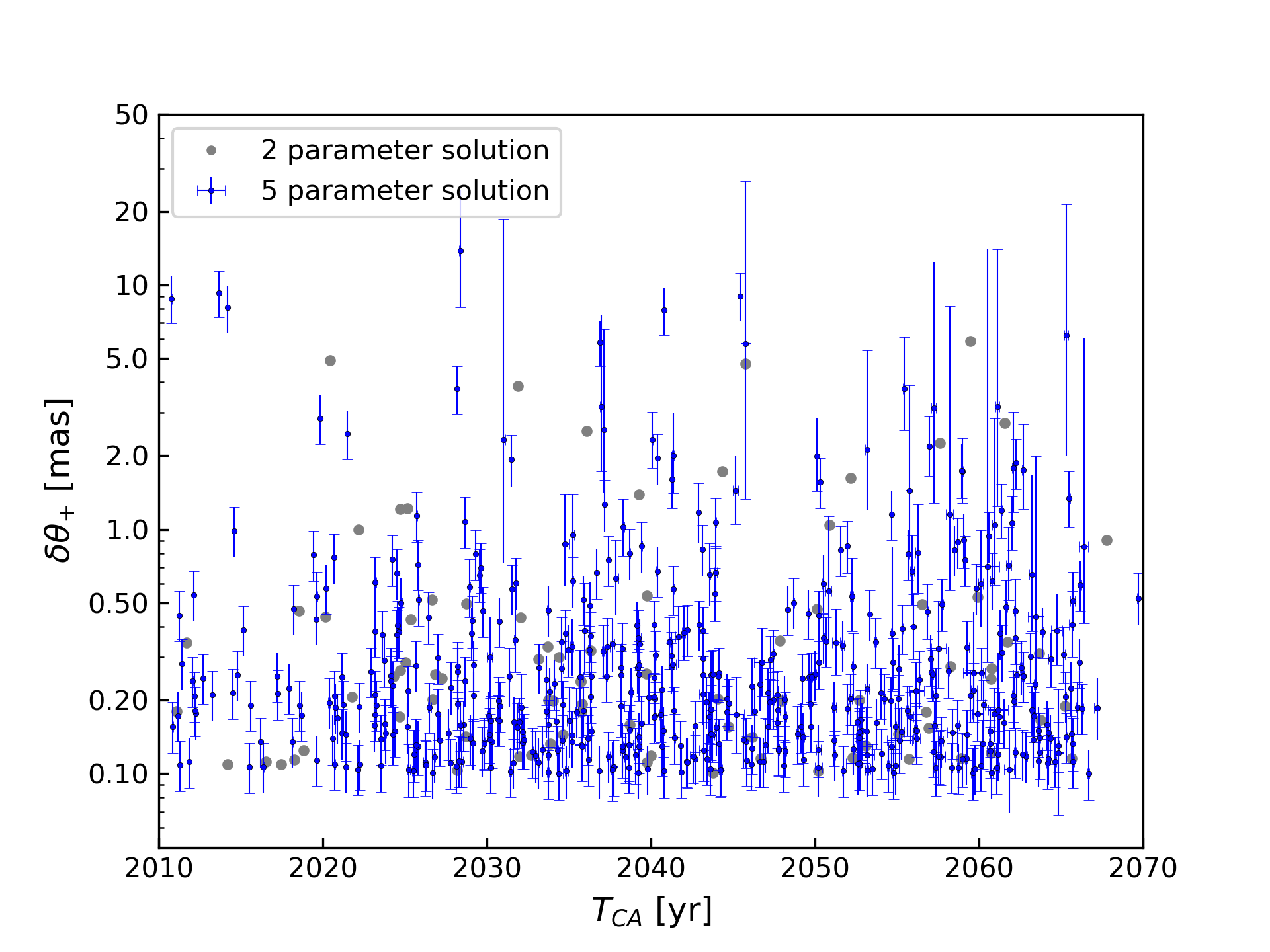}
\caption{Maximum shifts for all expected events caused by a WD, with (blue) and without (grey) five-parameter solution for the background source in \eDR.}
\label{Fig:WD} 
\end{figure}
For 625 (563, 409) events, the HPMS is classified as WD.
In the next decade 112 (95, 75)  events occur, out of which 9 (5, 4) have an expected angular shift larger than \(1\mas\).
As median we find an occurrence rate of 12 (10, 8) events per year with expected shifts larger than \(0.1\mas\), and about 1~event per year with expected shift larger than \(1\mas\), respectively (cf. Figure~\ref{Fig:WD}).
These events are of special interest, for two reasons. 
First, WDs are relatively faint for a given mass, i.e. the magnitude differences between lens and source are much smaller than for main sequence star-lenses. Hence, they provide ideal targets with large effects and only little blending.  
This is also shown by the larger fraction of events with smaller magnitude differences (\(66\%\) events with \(\Delta G<3\), compared to \(25\%\) for the full sample). 
Further, relations concerning the masses of WD are currently only poorly known, due to the lack of WD with directly determined masses. This increases the importance of direct mass measurements for  white dwarfs.

\cite{2018MNRAS.475...79H} estimates the event rate of astrometric microlensing events caused by stellar remnants. For \DRtwo sources and white dwarfs as lenses, they estimates \(1.52\cdot10^{-3}\) events per white dwarf per decade. With about 3800 white dwarfs in our sample of good HPMS, we would expect about 20 microlensing events between 2030 and 2065.5 (where our sample shows a constant behaviour). However, for the same time range, we found 146 events caused by a white dwarfs, with an expected shift larger than \(0.3\mas\) (the detection limit in \cite{2018MNRAS.475...79H}). This would lead to an event rate of \(1.1\cdot10^{-2}\). The higher event rate is due to our selection of HPMS only, which are more likely to cause astrometric microlensing events than the average population of white dwarfs.

\subsubsection{Events caused by binary stars}
Our sample also contains pairs of events with identical BGS, but different HPMS. In total, we found 24 such pairs, caused by 15 binary systems. A visual inspection showed that all of these are true events. 
For 15  of the 48  events the magnitude difference is below~\(6\).
Additionally, for 421 events, the HPMS is part of a binary system, but a measurable effect is not expected for the second component. 
We note that the orbital motion of the binary system is not included in our analysis, nor accounted for in the error estimation. It is possible that additional events are caused by binary stars which have angular separations larger than 30" (which we used as radius for a cross-match), or by binaries which are not resolved in \eDR, or where the other component does not satisfy our quality criteria.
Due to the orbital motion of the binary components, the true epoch and the separation of the closest approach might differ from our predictions. Corresponding corrections can be included in our analysis, once the binary motion is known. For some of the events, this will come with the full \Gaia Data Release 3 (i.e.~in 2022).

\subsubsection{Astrometric-microlensing events with photometric signature}
\begin{figure}
\center
\includegraphics[width=0.95\linewidth]{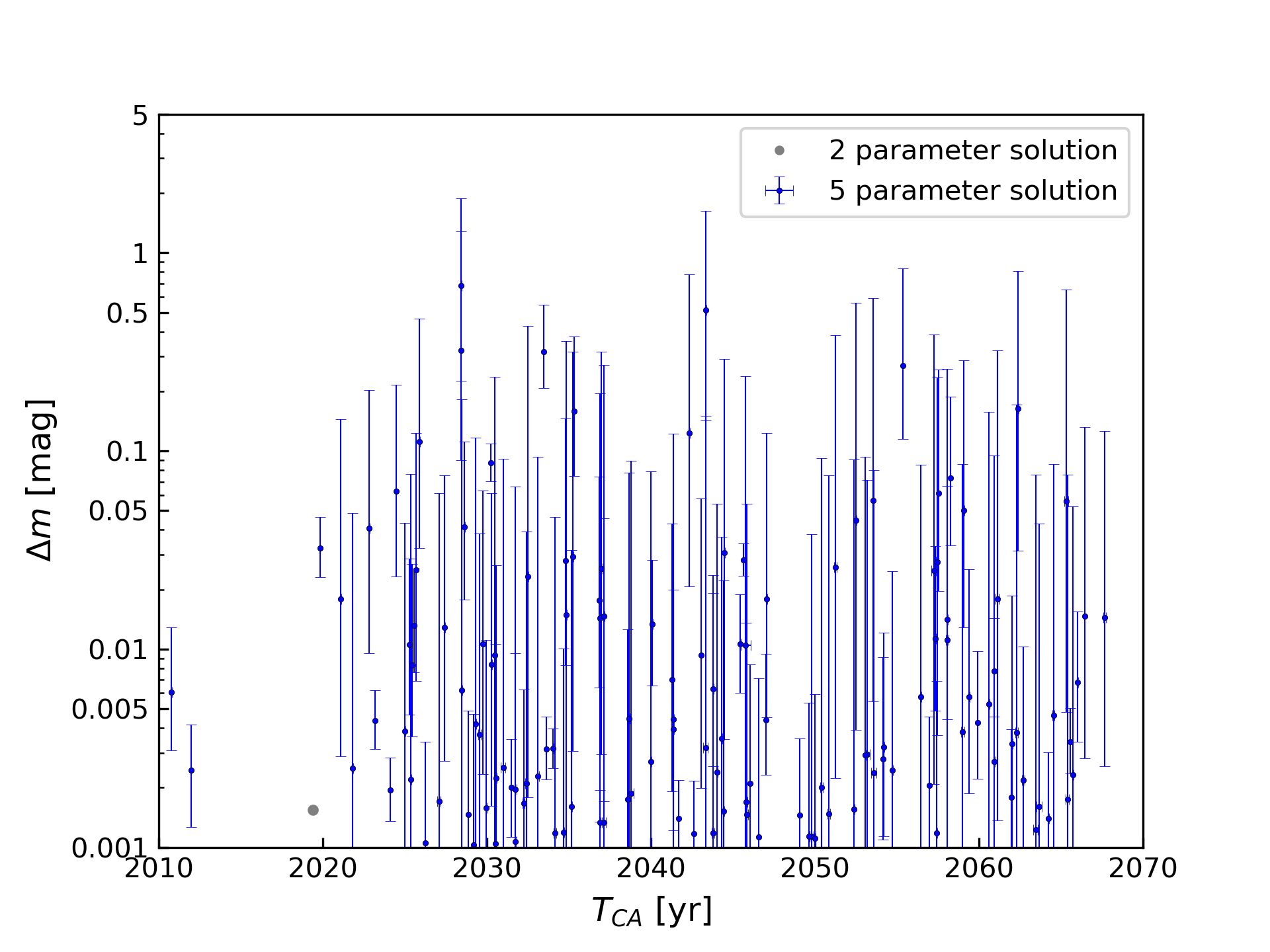}
\caption{Expected maximum magnification for all astrometric-microlensing events with photometric signature, except for one (grey dot) all events have a five-parameter solution for the background source in \eDR.
}
\label{Fig:phot} 
\end{figure}
In \Kt we predicted 127  astrometric-microlensing events which might lead to a noticeable photometric signal.
Using \eDR, we re-detect 51   of those events. Further 44  events were re-detected, but with an expected magnification below 0.1 mmag.
Additionally, we found 88 events with an expected magnification larger than \(1 \mmag\). 32  of those were predicted in \Kt but with an expected magnification below \(1\mmag\) the other newly predicted events are caused by stars with proper motions between \(100\mpy\) and \(150\mpy\). In total we found 139   astrometric-microlensing events with potential photometric signatures. However, all predicted magnifications have large uncertainties.
These are shown in Fig.~\ref{Fig:phot}. 
For  49 (49, 40)  events we expected a magnification larger than \(10\mmag\), and for  9 (9, 9)  events even larger than \(100\mmag\). We do not find any photometric events with high contrast (\(\Delta G> 6\)), due to the blending of the lens.  We note that our estimation only considers blending caused by the lens star, and neglects any blends from additional background sources.

\subsubsection{Astrometric-microlensing event by \Gaia eDR3-4053455379420641152}
\begin{figure}
\center
\includegraphics[width=0.95\linewidth]{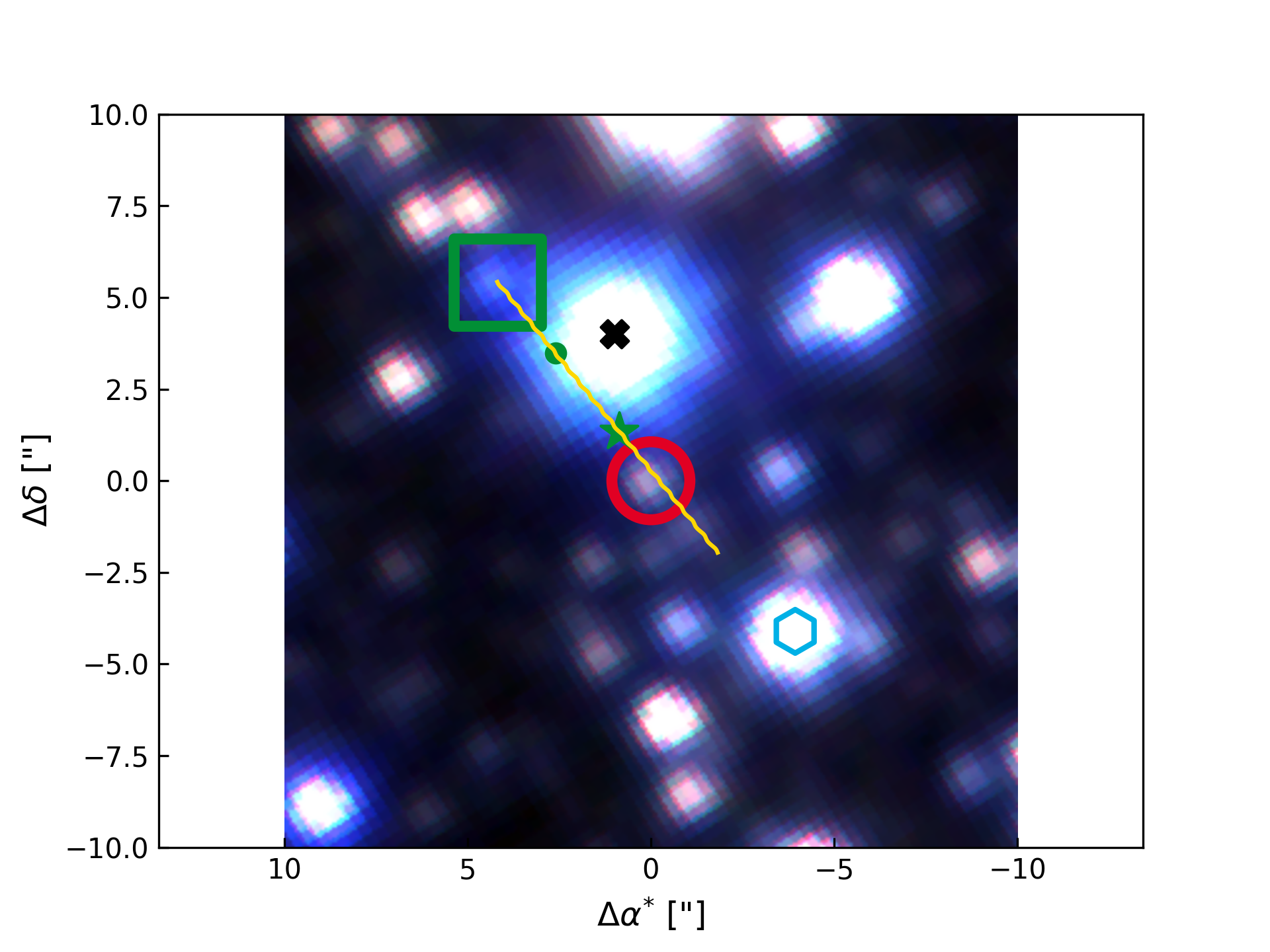}
\caption{H-J-Y colored cutout from VVV. centered on the position of the background source (red circle) The high-proper-motion lens star is indicated as by the green square (November 2010). Its motion between 2010 and 2030 is indicated by the yellow line, and its J2016.0, and J2021.5 position are given by the green dot and star, respectively. The small wobbling is due to its parallax. 
The binary companion is marked by the black cross.
The background star for the close encounter in 2036 is marked by the cyan hexagon.}
\label{Fig:VVV} 
\end{figure}

As an example, we here want to show a promising newly predicted astrometric-microlensing event caused by \Gaia eDR3-4053455379420641152 (VVV 176144893)  in 2025 (hereafter G-2025).
 G-2025 is located towards the galactic center(\({\rm RA}_{J2016} = 17^{\mathrm{h}}38^{\mathrm{m}}37^{\mathrm{s}}.04721\); \({\rm DEC}_{J2016} = -34^{\circ}27^{\prime}35^{\prime\prime}.4946\)) 
 with a proper motion of (\(\mu_\alpha*\),\,\(\mu_\delta\)) = (\(-316.1\mpy,\,-388.6\mpy)\), and a parallax of \(\varpi = 25.5\mas\). 
 \Gaia provides a magnitude of \(G = 17.8\MAG\), \( G_{BP} = 18.2 \MAG \) and \(G_{RP} = 16.9 \MAG\), respectively. 
 Therefore, we classify the star as a white dwarf. For those, we determined an approximated mass of \(m = 0.65 \Msun\).
G-2025 is located within the VVV footprint, and VVV lists a brightness of \(KS = 15.5 \MAG\).
A stamp from VVV is shown in Figure~\ref{Fig:VVV}, where G-2025 is marked by the green square. 
G-2025 has a similar proper motion  (\(\mu_\alpha^{*} = -326.3 \mpy\), \(\mu_\delta = -391.1\mpy\)) 
and parallax (\(\varpi = 25.9\mas\)) as the close-by star \Gaia eDR3-4053455379465036800 (2MASS J17383723-3427304; VVV 176850871, marked by the black cross in Fig.~\ref{Fig:VVV}).
This is also true for the proper motion provided by VIRAC \citep{2018MNRAS.474.1826S}. The motion can also be seen by comparing VVV images from different epochs. Hence, G-2025 and \Gaia eDR3-4053455379465036800 are forming a binary (or multiple stars) system.

In \(2025.17\pm 0.12\) it will pass by a \(G=20.25\MAG\) (\Gaia eDR3-4053455379420641152; VVV 176850869) background star with an impact parameter of \(d_{min} = (111\pm 69)\mas\). 
The large uncertainties are due to the missing five parameter solution for the background star in \eDR. 
Hence we used the displacement between \eDR and DR2 for the proper motion.
for the proper motion. We note that the so derived proper motion does not agree with the proper motion provided by VVV. Using the the VVV proper motion will lead to a \({\sim}10\%\) smaller minimal angular separation and thus to a  \({\sim}10\%\) larger shift, larger shift, but with an epoch of the closest approach about 6 months earlier (We can find large discrepancies between Gaia and VVV proper motion also for other stars close-by stars). 
The fourth \Gaia data release might provide proper motion and parallax for this source, which will help in the analysis after a successful observation of the event.
In Figure~\ref{Fig:VVV}, the background star is indicated by the red circle. 
For the background star VVV list J, H and Ks magnitudes of \(J=15.7\MAG\), \(H =14.8\MAG\) and \(Ks=14.4\MAG\), respectively. 

\begin{figure}
\center
\includegraphics[width=0.95\linewidth]{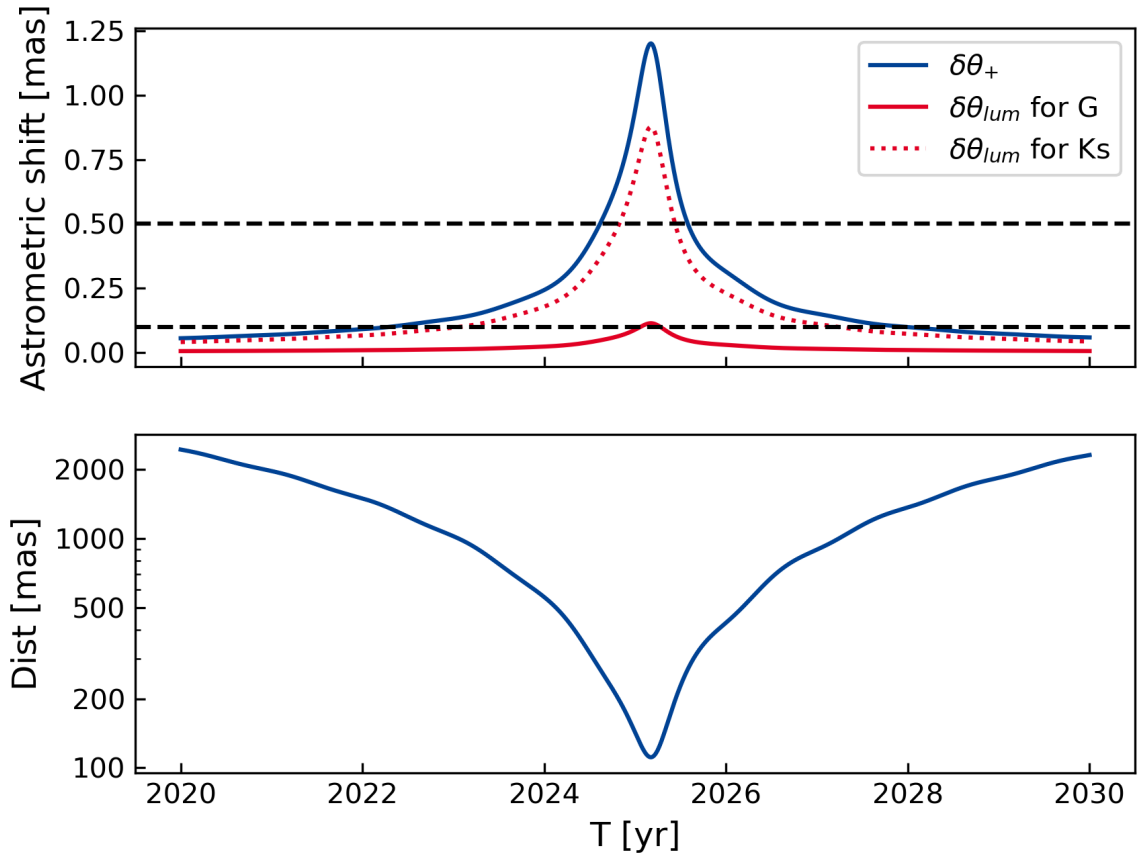}
\caption{Top: Expected shift of the major image (blue line), for the combined center of light using the G flux ratio (red solid line), and for the combined center of light using the Ks flux ratio (red dotted line).  The \(0.1\mas\) and \(0.5\mas\) level is indicated by the dashed lines. Bottom: angular separation between lens and source.}
\label{Fig:time} 
\end{figure}

We determined an Einstein radius of \(\theta_{E} = (11.6\pm1.5)\mas\). This results in an expected maximum shift of the major image of \(\delta\theta_{+} = 1.2^{+2.0}_{-0.5}\mas\) 
If it is not possible to resolve lens and source, only the combined center of light can be observed. For G-Band observation, we then expect a shift of \(\delta\theta_{lum} =0.11^{0.18}_{-0.05}\mas\).
Since in the near infrared, the source is brighter than the lens, a larger shift of the center of light can be observed. Using the Ks filter a maximum shift of
 \(\delta\theta_{lum, Ks} \simeq 0.87\mas\) is expected (red dotted line in Fig.~\ref{Fig:time}). JWST\footnote{James Webb Space Telescope, \url{https://www.jwst.nasa.gov}} will be able to measure this small shift. 
Both, the shift of the major image and the shift of the centre of light, as well as the angular separation between lens and source star, is shown in Figure ~\ref{Fig:time}as a function of time. The expected shift for the major image will be above \(0.5\mas\) over a period of about 11 months (about  5.5 years for \(\delta\theta_{+} > 0.1\mas\)). 

We found that G-2025 will pass by to further BGS. 
One in early 2028, with an expected impact parameter of\(d_{min} = (1300\pm60)\mas\).
For this event, we expect a shift of the major image of \(\delta\theta_{+} =(0.10\pm0.03)\mas\), which is only slightly above our selection criterion.
And one in 2036, with an expected impact parameter of \(d_{min} = (260\pm3)\mas\). This will lead to a shift of the major image of  \(\delta\theta_{+} = 0.52\pm0.13\mas\).
The BGS for the later event is marked by a cyan hexagon in Fig.~\ref{Fig:VVV}.

\begin{table*}
    \caption{Number of predicted astrometric-microlensing events:  all  events, events which lead to an shift including blending by a luminous lens, events with a White Dwarf as lens, and potential photometric microlensing signature. For each subset, three different ranges for the expected effect are given, and the first line is the sum of the three lines below. The number are also given for different magnitude differences between lens and source, where the last column gives the total number of events}
    \label{tab:result}
    \centering
    \begin{tabular}{r||c|c|c||c}
    & \multicolumn{4}{c}{number of events with}\\
	& \(\Delta G \le 3\) & \(3< \Delta G \le6\) & \(\Delta G>6\) & total \\
\hline\hline
\multicolumn{1}{l||}{all events} &&&&\\         
\(\delta \theta_{+} >= 0.1 \mas\) & 1246 & 1436 & 2160  & 4842\\
\(0.5 \mas > \delta \theta_{+} >= 0.1 \mas\)  & 1000 & 1121 & 1716 & 3837\\
\(1\mas > \delta \theta_{+}  > =0.5 \mas\)& 135 & 166 & 231 & 532\\
\(\delta \theta_{+}   >= 1 \mas\) &   111 & 149 & 213 & 473 \\
\hline
\multicolumn{1}{l||}{luminous lens}&&&&\\
\(\delta \theta_{lum}>= 0.1 \mas\) & 380 & 13 & 0 & 393 \\
\(0.5  \mas >\delta \theta_{lum}>= 0.1 \mas\) & 330 & 13 & 0 & 343 \\
 \(1 \mas > \delta \theta_{lum}>=0.5 \mas\) 35 & 0 & 0 & 35 \\
\(\delta \theta_{lum} >=1 \mas\) &  15 & 0 & 0 & 15 \\
\hline

\multicolumn{1}{l||}{events by White Dwarfs}&&&&\\
\(\delta \theta_{+} >= 0.1 \mas\) & 409 & 154 & 62 & 625 \\
\(0.5 \mas\ > \delta \theta_{+} >= 0.1 \mas\) & 333 & 129  & 43 & 505 \\
\(1\mas > \delta \theta_{+}  >= 0.5 \mas\) & 41 & 11 & 8 & 60 \\
\(\delta \theta_{+}   >= 1 \mas\) &  35 & 14 & 11 & 60 \\
\hline
\multicolumn{1}{l||}{photometric events} &&&&\\ 
\(\delta m>= 1 \mmag\) &   99 & 37 &  3 & 139 \\
\(10 \mmag > \delta m>= 1 \mmag\) &   59 &  28 &  3 & 90 \\
\(100 \mmag > \delta m>= 10 \mmag\) &   31 & 9 & 0  & 40 \\
\(\delta m>= 100 \mmag\) &  9 &  0 & 0 & 9 \\
\end{tabular}

\end{table*}

%

\section{Summary and Conclusion}
\label{Section:Conclusions}
From \eDR we found \(136,000\) close  passages of background stars by high proper motion (foreground) stars with proper motions larger than 100 mas per year. We defined a set of quality criteria to avoid spurious entries. Further, we sorted out potentially co-moving star pairs using the cross-match between \DRtwo and eDR3. 
By forecasting the motion of HPMS and BGS we derived the angular separations and epochs of the closest approach for each case. Using approximate masses for the lens stars, we estimated the expected microlensing effects, both astrometric and photometric. 

We give these predictions for 4842  events by 3791  distinct HPMS, where we expect a shift larger than \(0.1 \mas\). 3084  of these events had already been predicted from \DRtwo. For those we give updated parameters based on \eDR. For most of the events, we find a good agreement for the angular separation and epoch of the closest approach. Significant differences are mainly due to a significant changes of the proper motion values between  \DRtwo and eDR3. 

Compared to \Kt, we improved the exclusion of spurious sources and the detection of co-moving star pairs, especially in cases when \eDR does not provide a proper motion for the BGS. 
Due to the higher  precision in the proper motions, the uncertainties for the expected epochs and minimum separations decrease roughly by a factor of~2. 
However, the uncertainties of the expected shifts stay roughly constant since they are dominated by the errors of the assumed masses. This should be improved for future studies. However, measuring the masses is the ultimate aim of observing the astrometric-microlensing events. For that purpose the unlensed angular separation is of special interest, since it cannot directly be observed but is essential to determine the deflection. And this parameter does not depend on the assumed mass.

The observation of the predicted events and the subsequent mass determination will lead to a better understanding of stellar masses for isolated stars. 
Based on the prediction from \DRtwo, several observing programs started. With the present work, it is possible to  further improve the selection of promising events.
The 625  events caused by 473  different WD might be of particular interest. On the one hand, these are easier to observe, due to the lower magnitude difference (caused by the high mass-to-luminosity ratio). On the other hand, such observations will lead to a better understanding of WDs, and thus of the final phase of the evolution of low- to medium-mass stars.
It might also be valuable to observe multiple events caused by the same lens star. This can strongly reduce the uncertainties of the mass determination for such cases.

Using the upcoming \Gaia data releases will allow to further improve the analysis. 
Also, the future \Gaia releases will provide orbital parameters of binaries. These can strongly improve the selection of co-moving stars and might also be included in the forecast of the paths. 
Further improvements from future releases will be achieved due to the more precise astrometric parameters and due to the expected increase in the effective angular resolution. Finally, with the upcoming data releases, more detailed knowledge of the spurious sources is expected, which leads to better selections of BGS. New event predictions will be mainly added with epochs of closest approach during the \Gaia  mission or in the near future. 
These will be of special interest once the individual \Gaia measurements are published. 

Finally, we showed that the WD \Gaia eDR3-4053455379420641152 will cause a promising astrometric-microlensing event in 2025. This event is ideal for a JWST observation, since the background star is brighter in the near infrared wavelength regime, and a measurable shift of the center of light is expected even if it is not possible to resolve the lens and source stars.

\section*{Acknowledgements}
We would like to thank the referee who provided useful and detailed comments,  which helps to improve this paper.
We gratefully acknowledge the technical support we received from staff of the e-inf-astro project (BMBF F\"{o}rderkennzeichen 05A20VH5).\\
This work has made use of results from the ESA space mission \Gaia, the data from which were processed by the \Gaia Data Processing and Analysis Consortium (DPAC). Funding for the DPAC has been provided by national institutions, in particular the institutions participating in the \Gaia Multilateral Agreement. The \Gaia mission website is:
http://www.cosmos.esa.int/Gaia.\\
Based on data products from observations made with ESO Telescopes at the La Silla or Paranal Observatories under ESO programme ID 179.B-2002.\\
This publication makes use of data products from the Two Micron All Sky Survey, which is a joint project of the University of Massachusetts and the Infrared Processing and Analysis Center/California Institute of Technology, funded by the National Aeronautics and Space Administration and the National Science Foundation.\\
The Pan-STARRS1 Surveys (PS1) and the PS1 public science archive have been made possible through contributions by the Institute for Astronomy, the University of Hawaii, the Pan-STARRS Project Office, the Max-Planck Society and its participating institutes, the Max Planck Institute for Astronomy, Heidelberg and the Max Planck Institute for Extraterrestrial Physics, Garching, The Johns Hopkins University, Durham University, the University of Edinburgh, the Queen's University Belfast, the Harvard-Smithsonian Center for Astrophysics, the Las Cumbres Observatory Global Telescope Network Incorporated, the National Central University of Taiwan, the Space Telescope Science Institute, the National Aeronautics and Space Administration under Grant No. NNX08AR22G issued through the Planetary Science Division of the NASA Science Mission Directorate, the National Science Foundation Grant No. AST-1238877, the University of Maryland, Eotvos Lorand University (ELTE), the Los Alamos National Laboratory, and the Gordon and Betty Moore Foundation.\\
This research has made use of the SIMBAD database and the ``Aladin sky atlas" \citep{2000A&AS..143...33B} both developed and operated at CDS, Strasbourg Observatory, France

\software{
astropy \citep{astropy}, 
astroquery, \citep{astroquery}, 
matplotlib \citep{Hunter:2007},
numpy \citep{harris2020array},
TOPCAT \citep{2005ASPC..347...29T}, 
}
\bibliography{AML_edr3.bbl}

\begin{table*}
\caption{Description of the columns in the online table. 
This part list all the important parameters 
directly taken from \eDR. It is 
continued in Table\ref{tab:column_description_continue}), where all the from us derived columns are described.}
\footnotesize
\label{tab:column_description}

\begin{tabular}{r|c|c|c}

& Name & Unit& Description \\ 
\hline
1 &   event\_id &  &  \begin{tabular}{@{}c@{}}Unique identifier of this event. This is the decimal representation of the \\ Gaia eDR3 source\_id of the lens, and a disambiguator for the lensed object \end{tabular} \\
\hline
2 &   lens\_id &  & Gaia eDR3 source\_id of the lens \\
3 &   lens\_ra &  deg & Gaia eDR3 RA of the lens \\
4 &   lens\_dec &  deg & Gaia eDR3 Declination of the lens \\
5 &   lens\_err\_ra &  mas & Error of lens\_\_ra from Gaia eDR3 \\
6 &   lens\_err\_dec &  mas & Error of lens\_\_dec from Gaia eDR3 \\
7 &   lens\_pmra &  mas/yr &      Proper motion in RA of the lens from Gaia eDR3 \\
8 &   lens\_pmdec &  mas/yr &      Proper motion in Declination of the lens from Gaia eDR3 \\
9 &   lens\_err\_pmra &  mas/yr &      Error of lens\_pmra from Gaia eDR3 \\
10 &  lens\_err\_pmdec &  mas/yr &      Error of lens\_pmdec from Gaia eDR3 \\
11 &  lens\_parallax &  mas &     Parallax of the lens from Gaia eDR3 \\
12 &  lens\_err\_parallax &  mas &     Standard error of the parallax the lens from Gaia eDR3 \\
13 &  lens\_phot\_g\_mean\_mag &  mag &     Mean magnitude of the lens in the integrated G band from Gaia eDR3 \\
14 &  lens\_phot\_rp\_mean\_mag &  mag &     Mean magnitude of the lens in the integrated RP band from Gaia eDR3 \\
15 &  lens\_phot\_bp\_mean\_mag &  mag &     Mean magnitude of the lens in the integrated BP band from Gaia eDR3 \\
\hline
16 &  ob\_id & & Gaia eDR3 source\_id of the lensed object \\
17 &  ob\_ra &  deg &     Gaia eDR3 RA of the lensed object  \\
18 &  ob\_dec &  deg &     Gaia eDR3 Declination of the lensed object  \\
19 &  ob\_err\_ra &  mas &     Error of ob\_\_ra from Gaia eDR3 \\
20 &  ob\_err\_dec &  mas &     Error of ob\_\_dec from Gaia eDR3  \\
21 &  ob\_pmra &  mas/yr &      Proper motion in RA of the lensed object from Gaia eDR3 \\
22 &  ob\_pmdec &  mas/yr &     Proper motion in Declination of the lensed object from Gaia eDR3 \\
23 &  ob\_err\_pmra &  mas/yr & Error of ob\_pmra from Gaia eDR3 \\
24 &  ob\_err\_pmdec &  mas/yr & Error of ob\_pmdec from Gaia eDR3 \\
25 &  ob\_parallax &  mas &     Parallax of the lensed object from Gaia eDR3 \\
26 &  ob\_err\_parallax &  mas &     Standard error of the parallax the lensed object from Gaia eDR3 \\
27 &  ob\_phot\_g\_mean\_mag &  mag &     Mean magnitude of the lensed object in the integrated G band from Gaia eDR3 \\
28 &  ob\_phot\_rp\_mean\_mag &  mag &     Mean magnitude of the lensed object in the integrated RP band from Gaia eDR3 \\
29 &  ob\_phot\_bp\_mean\_mag &  mag &     Mean magnitude of the lensed object in the integrated BP band from Gaia eDR3 \\
30 &  ob\_displacement\_ra\_doubled &  mas &     Doubled displacement in RA between DR2 and DR3, cos(\(\delta\)) applied \\
31 &  ob\_displacement\_dec\_doubled &  mas &     Doubled displacement in Dec between DR2 and DR3 \\
\hline
\end{tabular}
\end{table*}
\begin{table*}
\caption{Continuation of Table\ref{tab:column_description}. Description of the columns estimated in this analysis}
\label{tab:column_description_continue}
\footnotesize
\begin{tabular}{r|c|c|c}
& Name & Unit& Description \\ 
\hline
32 & star\_type & &  \begin{tabular}{@{}c@{}}Type  of the lensing star:  \\ WD = White Dwarf, MS = Main Sequence, RG = Red Giant, BD = Brown Dwarf \end{tabular} \\
33 &  mass &  solMass &     Estimated mass of the lens from Gaia eDR3 \\
34 &  err\_mass &  solMass &     Error in the mass of the lens from Gaia eDR3 \\
35 &  theta\_e &  mas &     Einstein radius of the event \\
36 &  err\_theta\_e &  mas &     Error in the Einstein radius of the event \\
37 &  t\_aml &  yr & Approximate duration of the event (i.e., shift \(\gtrsim 0.1\mas\)) \\
37 &  t\_0\_1mas &  yr & \begin{tabular}{@{}c@{}}Approximate duration on which shift\_plus changes by 0.1 mas from its maximum \\( abs(shift\_plus\_vector(tca)-shift\_plus\_vector(tca-t\_0\_1\_mas)) = 0.1 mas ) \end{tabular}  \\
37 &  t\_50pc &  yr & \begin{tabular}{@{}c@{}}Approximate duration on which shift\_plus changes by \(50\%\) from its maximum \\( abs(shift\_plus\_vector(tca)-shift\_plus\_vector(tca-t\_50pc)) = 50\% * shift\_plus )\end{tabular}  \\
40 &  tca &  yr & Estimated time of the closest approach \\
41 &  err\_tca &  yr &  Error in tca \\
42 &  dist &  mas &     Estimated distance at closest approach \\
43 &  err\_dist &  mas &     Error d\_min \\
44 &  u &  & Estimated distance at closest approach in Einstein radii \\
45 &  u\_error\_m &  & Left 67\% confidence interval of u \\
46 &  u\_error\_p &  & Right 67\% confidence interval of u \\
47 &  shift &  mas &     Maximal astrometric shift of the center of light \\
48 &  shift\_error\_m &  mas &     Left 67\% confidence interval of shift \\
49 &  shift\_error\_p &  mas &     Right 67\% confidence interval of shift \\
50 &  shift\_lum &  mas &     Maximal astrometric shift including lens-luminosity effects \\
51 &  shift\_lum\_error\_m &  mas &     Left 67\% confidence interval of shift\_lum \\
52 &  shift\_lum\_error\_p &  mas &     Right 67\% confidence interval of shift\_lum \\
53 &  shift\_plus &  mas &     Maximal astrometric shift of brighter image \\
54 &  shift\_plus\_error\_m &  mas &     Left 67\% confidence interval of shift\_plus \\
55 &  shift\_plus\_error\_p &  mas &     Right 67\% confidence interval of shift\_plus \\
56 &  magnification &  mag &     Maximal magnification \\
57 &  magnification\_error\_m &  mag &     Left 67\% confidence interval of magnification \\
58 &  magnification\_error\_p &  mag &     Right 67\% confidence interval of magnification \\
\hline
59 &  l2\_tca &  yr & Estimated time of the closest approachas seen from Earth-Sun L2 point \\
60 &  l2\_err\_tca &  yr & Error in l2\_tca \\
61 &  l2\_dist &  mas &     Estimated distance at closest approachas seen from Earth-Sun L2 point \\
62 &  l2\_err\_dist &  mas &     Error l2\_d\_min \\
63 &  l2\_u & & Estimated distance at closest approach in Einstein radiias seen from Earth-Sun L2 point \\
64 &  l2\_u\_error\_m & & Left 67\% confidence interval of l2\_u \\
65 &  l2\_u\_error\_p & & Right 67\% confidence interval of l2\_u \\
66 &  l2\_shift &  mas &     Maximal astrometric shift of the center of lightas seen from Earth-Sun L2 point \\
67 &  l2\_shift\_error\_m &  mas &     Left 67\% confidence interval of l2\_shift \\
68 &  l2\_shift\_error\_p &  mas &     Right 67\% confidence interval of l2\_shift \\
69 &  l2\_shift\_lum &  mas &     Maximal astrometric shift including lens-luminosity effectsas seen from Earth-Sun L2 point \\
70 &  l2\_shift\_lum\_error\_m &  mas &     Left 67\% confidence interval of l2\_shift\_lum \\
71 &  l2\_shift\_lum\_error\_p &  mas &     Right 67\% confidence interval of l2\_shift\_lum \\
72 &  l2\_shift\_plus &  mas &     Maximal astrometric shift of brighter imageas seen from Earth-Sun L2 point \\
73 &  l2\_shift\_plus\_error\_m &  mas &     Left 67\% confidence interval of l2\_shift\_plus \\
74 &  l2\_shift\_plus\_error\_p &  mas &     Right 67\% confidence interval of l2\_shift\_plus \\
75 &  l2\_magnification &  mag &     Maximal magnificationas seen from Earth-Sun L2 point \\
76 &  l2\_magnification\_error\_m &  mag &     Left 67\% confidence interval of l2\_magnification \\
77 &  l2\_magnification\_error\_p &  mag &     Right 67\% confidence interval of l2\_magnification \\
\hline
\end{tabular}
\end{table*}

\end{document}